\documentclass[%
 reprint,
 amsmath,amssymb,
 aps,
floatfix,
]{revtex4-2}

\usepackage[colorlinks=true,
linkcolor=blue,
citecolor=blue,
urlcolor=blue]{hyperref}
\usepackage{xcolor}
\usepackage{breakurl}
\usepackage{url}
\usepackage{graphicx}
\usepackage{dcolumn}
\usepackage{bm}

\begin{document}

\preprint{APS/123-QED}
\title{ Non-Hermitian Purcell Physics in Dissipatively Coupled Planar Photon–Magnon Systems}

\author{Shubham Singh}
\author{Sachin Verma}
\author{Animesh Chakraworty}
\author{Abhishek Maurya}
\author{Biswanath Bhoi$^{*}$}
\author{Rajeev Singh}
 \email{Corresponding author, E-mail: biswanath.phy@itbhu.ac.in,       rajeevs.phy@itbhu.ac.in}
\affiliation
 {Department of Physics, Indian Institute of Technology (Banaras Hindu University) Varanasi, Varanasi - 221005, India.}


\begin{abstract}
The Purcell effect has emerged as a powerful mechanism for controlling spontaneous emission and dissipation in cavity and nanophotonic systems; however, its realization in dissipative non-Hermitian hybrid platforms remains largely unexplored. In this work, we investigate the Purcell effect in a dissipatively coupled photon–magnon hybrid quantum system consisting of a yttrium iron garnet (YIG) thin film integrated with an inverted octa-ring resonator (IORR) in a planar geometry. To describe the underlying dissipative hybrid dynamics, we develop a quantum theoretical framework based on non-Hermitian coupled-mode theory combined with the input–output formalism also we analyze the temporal decay dynamics of the Photon–Magnon System. Full-wave electromagnetic simulations demonstrate that the system can be engineered to operate in the level-attraction regime through dissipative photon–magnon coupling. By systematically tuning the magnon damping, we uncover the emergence of a Purcell regime in which cavity-photon dissipation is selectively enhanced through magnon-mediated loss channels. We further show that the saturation magnetization ($M_s$) provides an additional degree of control over the onset, tunability, and robustness of the Purcell enhancement. The combined modulation of magnon damping and $M_s$ strongly influences the effective photon–magnon coupling, hybrid-mode evolution, and dissipation landscape, enabling precise control of hybrid quantum states. These findings establish a versatile strategy for engineering magnetization and dissipation-controlled photon–magnon interactions in planar, chip-compatible architectures, opening new avenues for non-Hermitian cavity magnonics, tunable microwave dissipation engineering, and hybrid quantum information technologies.
\end{abstract}

\maketitle


\section{\label{sec:level1}INTRODUCTION }

Open quantum systems continuously exchange energy and information with their surrounding environment, giving rise to dissipation, decoherence, and irreversible dynamics that cannot be described within the framework of conventional Hermitian quantum mechanics\cite{Rao2019, Gollwitzer2021, Kurizki2015, LachanceQuirion2019, Zhang2023}. The interaction between a system and its environment introduces effective non-Hermitian behavior, where gain and loss fundamentally modify both the eigenvalue spectrum and dynamical evolution \cite{ZareRameshti2022, Boventer2019}. Such systems exhibit a variety of unconventional phenomena, including exceptional points \cite{Junyoung2026, Xu2016}, level attraction \cite{Liu2020}, linewidth bifurcation \cite{Mi2025}, mode coalescence\cite{Nair2022} , and topology-dependent state evolution\cite{Han2024, Harder2018}. Over the past decade, these unique characteristics have stimulated extensive research across photonic \cite{Mahboobeh2024}, optomechanical \cite{Chen2023, Aspelmeyer2014}, superconducting \cite{Tabuchi2016, Morris2017}, and spin-based quantum platforms \cite{Atature2018, Banerjee2025}, where engineered dissipation has emerged as a powerful resource for manipulating energy transfer, wave propagation, and light–matter interactions \cite{Metelmann2015}.

\begin{figure*}[t]
    \centering
    \includegraphics[width=\textwidth]{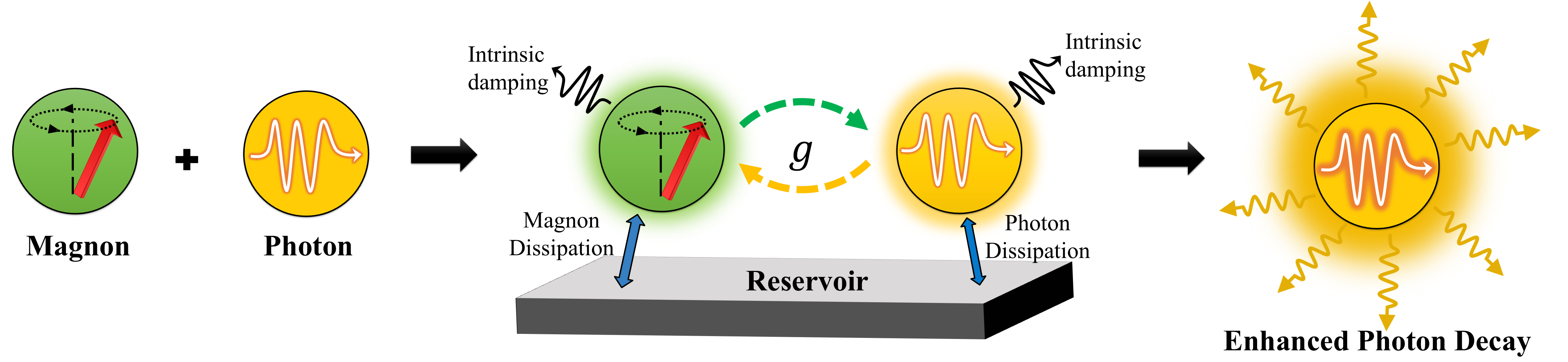}
    \caption{Schematic illustration of the Purcell effect in a dissipatively coupled magnon--photon hybrid system.}
    \label{fig:1}
\end{figure*}

Among the various open quantum platforms, cavity magnonics has become an attractive system for exploring coherent and dissipative interactions between microwave photons and collective spin excitations (magnons) in magnetic materials \cite{Harder2021, Shuai2025, Maurya2024}. Owing to their long coherence times, magnetic-field tunability, and compatibility with microwave quantum technologies, photon–magnon hybrid systems have been widely investigated for coherent information transfer \cite{Yuan2022}, quantum transduction \cite{Wang2022}, storage application\cite{Heshami2016, Zhang2015}, microwave signal processing \cite{Li2020}, sensing \cite{Yang2024}, and nonreciprocal devices \cite{Shuai2025}. Unlike coherent coupling, which produces level repulsion, dissipative coupling mediated by a common environment leads to level attraction, linewidth evolution, and mode coalescence, providing a hallmark of non-Hermitian cavity magnonics \cite{Wang2020, Bhoi2019}. These phenomena enable dissipation engineering for controlling spectral topology and energy transport. Consequently, non-Hermitian cavity magnonics has emerged as a powerful platform for exploring exceptional-point physics, topological mode evolution, and reservoir-engineered hybrid dynamics \cite{Junyoung2026}. However, most studies have focused on spectral signatures such as level attraction and exceptional points, while the influence of dissipative coupling on hybrid-mode stability and energy dissipation remains largely unexplored. In particular, the use of engineered magnon losses as a controllable dissipative reservoir for tailoring cavity-photon decay has received little attention, especially in planar, chip-compatible cavity-magnonic architectures.

An important consequence of reservoir engineering in open quantum systems is the Purcell effect, whereby coupling to an auxiliary resonant mode modifies the decay rate of an emitter or resonator. Although originally formulated in cavity quantum electrodynamics as the enhancement of spontaneous emission \cite{Stanfield2023, Auffeves2007, Krasnok2015, Kaupp2016}, the underlying concept extends naturally to hybrid photon–magnon systems, where cavity photons can acquire additional decay channels through their interaction with lossy magnon modes (Fig.\ref{fig:1}). Recent studies have demonstrated Purcell-enhanced photon decay in cavity magnonic systems by increasing magnon dissipation \cite{Zhao2023, Zhao2023APL, Verma2026}. Nevertheless, the interplay between dissipative photon–magnon coupling (PMC), non-Hermitian spectral evolution, and Purcell-enhanced cavity decay has not yet been systematically established. In particular, it remains unclear how magnetic damping and intrinsic magnetic properties govern the transition from strong hybridization to a dissipation-dominated Purcell regime in a hybrid photon–magnon systems.

Motivated by these open questions, we investigate the emergence of Purcell-enhanced photon dissipation in a dissipatively coupled non-Hermitian photon–magnon system consisting of a yttrium iron garnet (YIG) thin film coupled to an inverted octagon-ring resonator (IORR). The hybrid structure is first studied using full-wave electromagnetic simulations in CST Microwave Studio, where the magnon damping parameter ($\alpha$) and saturation magnetization ($M_s$) are varied to examine the evolution of the coupled photon–magnon modes. To interpret and validate the simulation results, we develop a theoretical model based on non-Hermitian coupled-mode theory and the input–output formalism, incorporating both coherent and dissipative photon–magnon interactions. From this model, we derive the condition for the Purcell regime in terms of the mode damping rates, dissipative coupling strength, and detuning. Applying this criterion to the simulated system allows us to identify the parameter regime in which enhanced photon dissipation occurs. The transition into this regime is accompanied by level attraction, hybrid-mode coalescence, and linewidth broadening, indicating the increasing influence of the dissipative magnonic mode on the cavity photon. More importantly, we show that Purcell-enhanced photon decay emerges naturally as a consequence of non-Hermitian reservoir engineering mediated by dissipative PMC rather than as an independent phenomenon. This work establishes a unified relationship between dissipative coupling, spectral topology, cavity dissipation, and Purcell dynamics in planar cavity magnonics, providing new physical insights into non-Hermitian hybrid systems and offering practical design principles for dissipation-engineered microwave photonic, magnonic, and chip-scale quantum technologies. 
 
\section{NUMERICAL MODELING AND SIMULATION SETUP}
\subsection{Simulation Geometry and Open-System Configuration}

\begin{figure}
    \centering
    \includegraphics[width=\columnwidth]{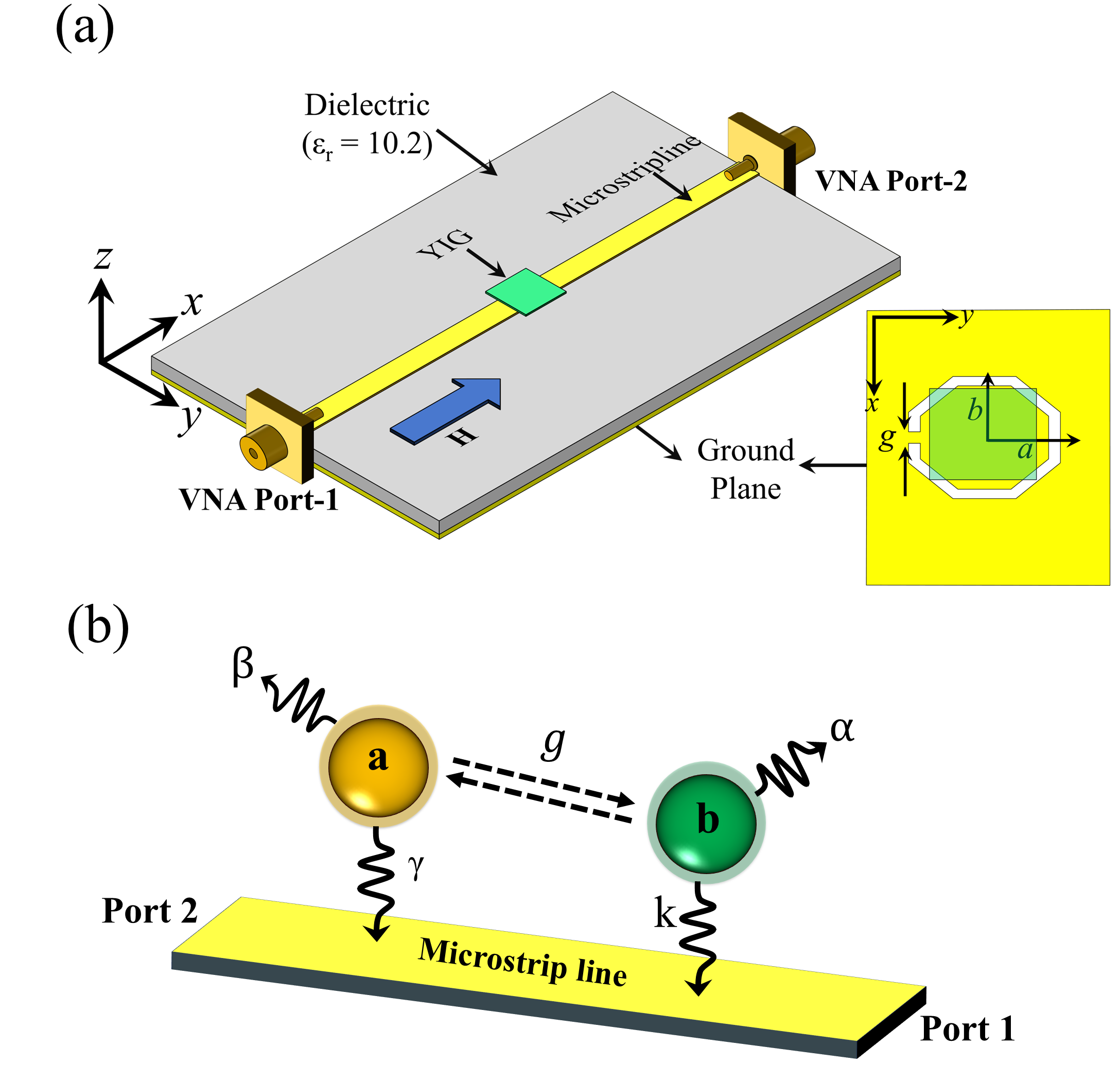}
    \caption{The simulation setup is shown in (a) hybrid system composed of a YIG film (green) and a microstrip line (yellow) fabricated on an RO3010 substrate with a dielectric constant of 10.2. The substrate thickness is 0.6 mm, and the ground plane thickness is 35 µm. The inverted octagonal-ring resonator (IORR) has an outer radius of a=3mm, an inner radius of b=2mm, and a gap of g =0.64mm. The YIG film dimensions are 4mm x 4mm x 20 µm , while the microstrip line has a width of 0.57 mm and a thickness of 35 µm. (b) Schematic diagram of the quantum coupled-mode model. }
    \label{fig:2}
\end{figure}

To explore the possibility of Purcell-enhanced photon dissipation, we investigate a hybrid photon–magnon system in which the coupled modes interact with a common electromagnetic environment. A schematic of the simulated structure is shown in Fig.~\ref{fig:2}\textcolor{blue}{(a)}. The system consists of a planar inverted octagon-ring resonator (IORR) integrated with a yttrium iron garnet (YIG) thin film and coupled to a microstrip transmission line. The microstrip line serves not only as the excitation and readout channel but also as an effective electromagnetic reservoir for the hybrid system. Because both the resonator photon mode and the magnon mode in the YIG film can exchange energy with this common environment, the system is intrinsically open and its dynamics cannot be described solely by a Hermitian Hamiltonian \cite{Yu2019}. In particular, the radiative leakage of the photon mode into the transmission line provides a dissipative decay channel, while the coupling of the YIG magnetization to the microwave electromagnetic field allows the magnon excitation to dissipate through the same electromagnetic environment. The coexistence of these decay pathways provides a physical basis for environment-mediated dissipative coupling between the photon and magnon modes \cite{Harder2021}. The IORR geometry is particularly suitable for this purpose because it provides a localized microwave magnetic field near its inner edges, resulting in appreciable spatial overlap with the YIG film and enabling photon–magnon interaction. At the same time, its open geometry permits coupling of the resonator field to the surrounding electromagnetic environment through radiative leakage. Thus, the structure provides a planar electromagnetic realization of an open photon–magnon system in which both coherent hybridization and dissipation can be controlled through the resonator and magnetic properties.

To investigate the resulting open-system dynamics, full-wave electromagnetic simulations are performed using the frequency-domain solver of CST Microwave Studio. The microwave response is characterized by the forward scattering parameter ($|S_{21}|$) as a function of excitation frequency and externally applied static magnetic field $H_{dc}$. The magnetic field is applied along the x-axis, allowing the ferromagnetic resonance frequency (FMR) of the YIG film to be continuously tuned and, consequently, the detuning between the magnon and photon modes to be controlled. The YIG film is modeled using experimentally relevant magnetic parameters, including the gyromagnetic ratio $\gamma =1.76\times {10}^{11}\; rad\cdot T^{-1}s^{-1} $ and saturation magnetization ($M_s = 0.175$ T), while magnetocrystalline anisotropy is neglected. The geometric dimensions of the IORR, microstrip transmission line, and YIG film are provided in the caption of Fig.~\ref{fig:2}\textcolor{blue}{(a)}. 

To systematically examine the role of magnetic dissipation, the intrinsic Gilbert damping constant ($\alpha$) of the YIG film is varied over the range $1.4\times {10}^{-4}$  to $2.8\times {10}^{-2}$. In the simulations, varying $\alpha$ directly modifies the FMR linewidth $\Delta H$ and hence the magnon dissipation rate. This allows the hybrid-mode evolution to be tracked as the system is driven from a low-loss strong-coupling regime towards a dissipation-dominated regime. In addition, the saturation magnetization\cite{Verma2025b} is varied over $M_{s}$ = 90 - 175 mT to examine its effect on the magnon resonance and photon–magnon interaction. The combined variation of ($\alpha$) and ($M_{s}$) provides control over the magnon resonance and dissipation, enabling us to examine their influence on the hybrid-mode spectra, linewidth evolution, and photon decay, and to identify the parameter regime associated with Purcell-enhanced photon dissipation.

\subsection{Effect of Magnon Damping on Hybrid-Mode Evolution}

\begin{figure}
    \centering
    \includegraphics[width=\columnwidth]{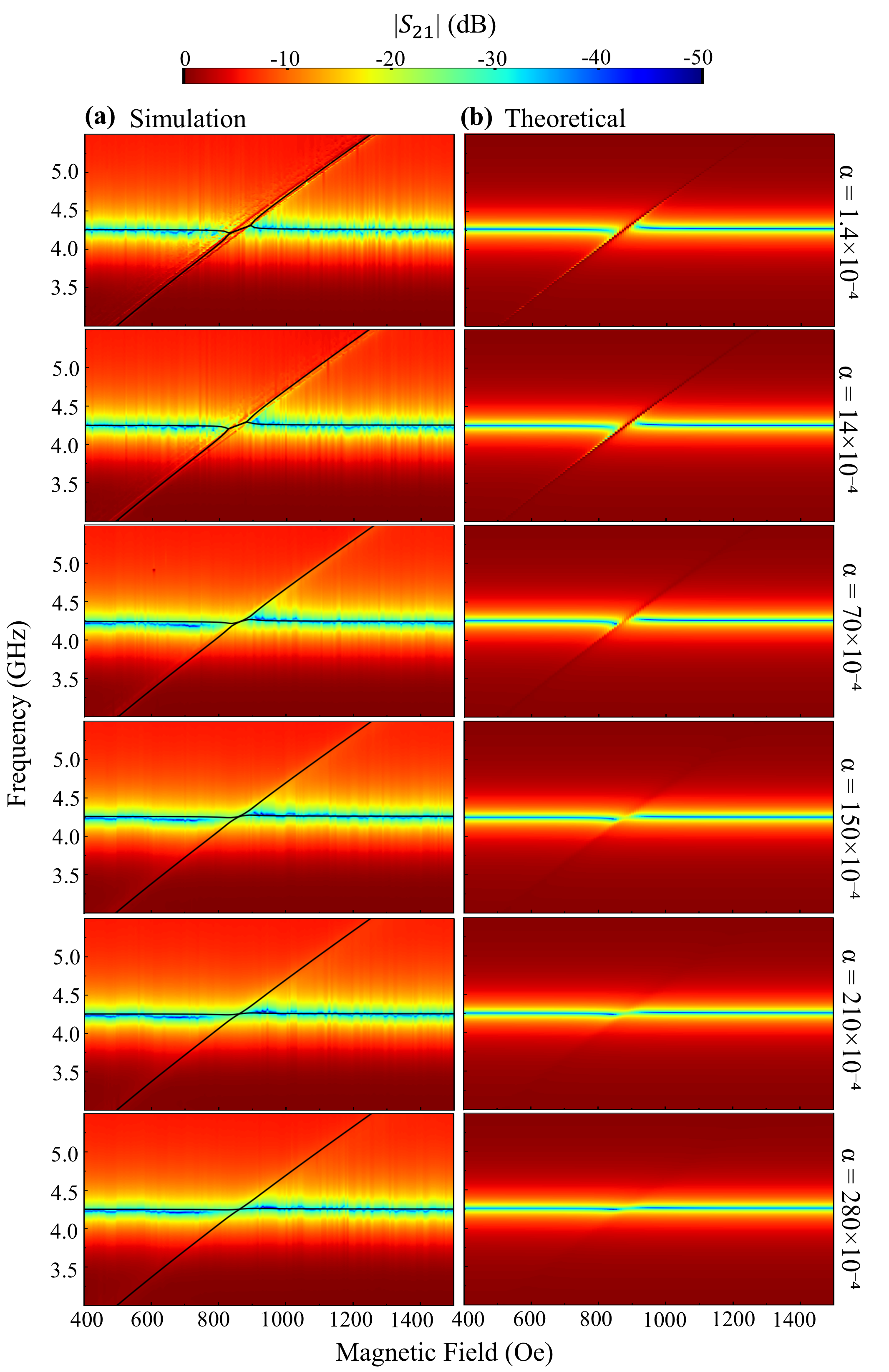}
    \caption{(a) Simulated and (b) Theoretical  microwave transmission spectra $|S_{21}|$ (0 to -50 dB) in the frequency–magnetic field (f–H) plane for $M_s$=175mT and different magnon damping values. The right panel shows the simulated spectra with fitted eigenvalue solutions (black solid lines), while the left panel presents the corresponding theoretical spectra. }
    \label{fig:3}
\end{figure}

Figure ~\ref{fig:3}\textcolor{blue}{(a)} presents the CST-simulated transmission spectra, $S_{21}$ as a function of excitation frequency and applied magnetic field for different values of the intrinsic magnon damping parameter $\alpha$. For relatively low damping, two hybrid-mode branches are observed, with the branches approaching each other near the photon–magnon resonance. As $\alpha$ increases, the frequency merging region becomes progressively narrower, while the resonance linewidths broaden and the transmission amplitude decreases. At sufficiently large damping, the hybrid resonances become strongly broadened and the transmission is substantially suppressed, indicating the increasing influence of magnon dissipation on the hybrid-system response\cite{Zhao2023APL}.

The observed linewidth broadening can be attributed primarily to the increased magnetic loss, since the intrinsic cavity loss remains essentially unchanged in the simulations. Through photon–magnon hybridization, energy initially stored in the cavity mode can be transferred to the magnon mode. When the magnon damping is increased, this transferred energy is dissipated more rapidly into the environment, thereby reducing the energy that can be coherently exchanged back to the cavity. As a result, the photon-like hybrid mode acquires an additional loss channel and its effective linewidth increases with $\alpha$. The damping-dependent evolution of the spectra therefore provides a direct numerical indication of dissipation-induced enhancement of photon decay. The detailed connection between this enhanced decay and the Purcell effect is established later using the theoretical open-system model.

\subsection{Effect of Saturation Magnetization on Photon–Magnon Interaction}

\begin{figure}
    \centering
    \includegraphics[width=\columnwidth]{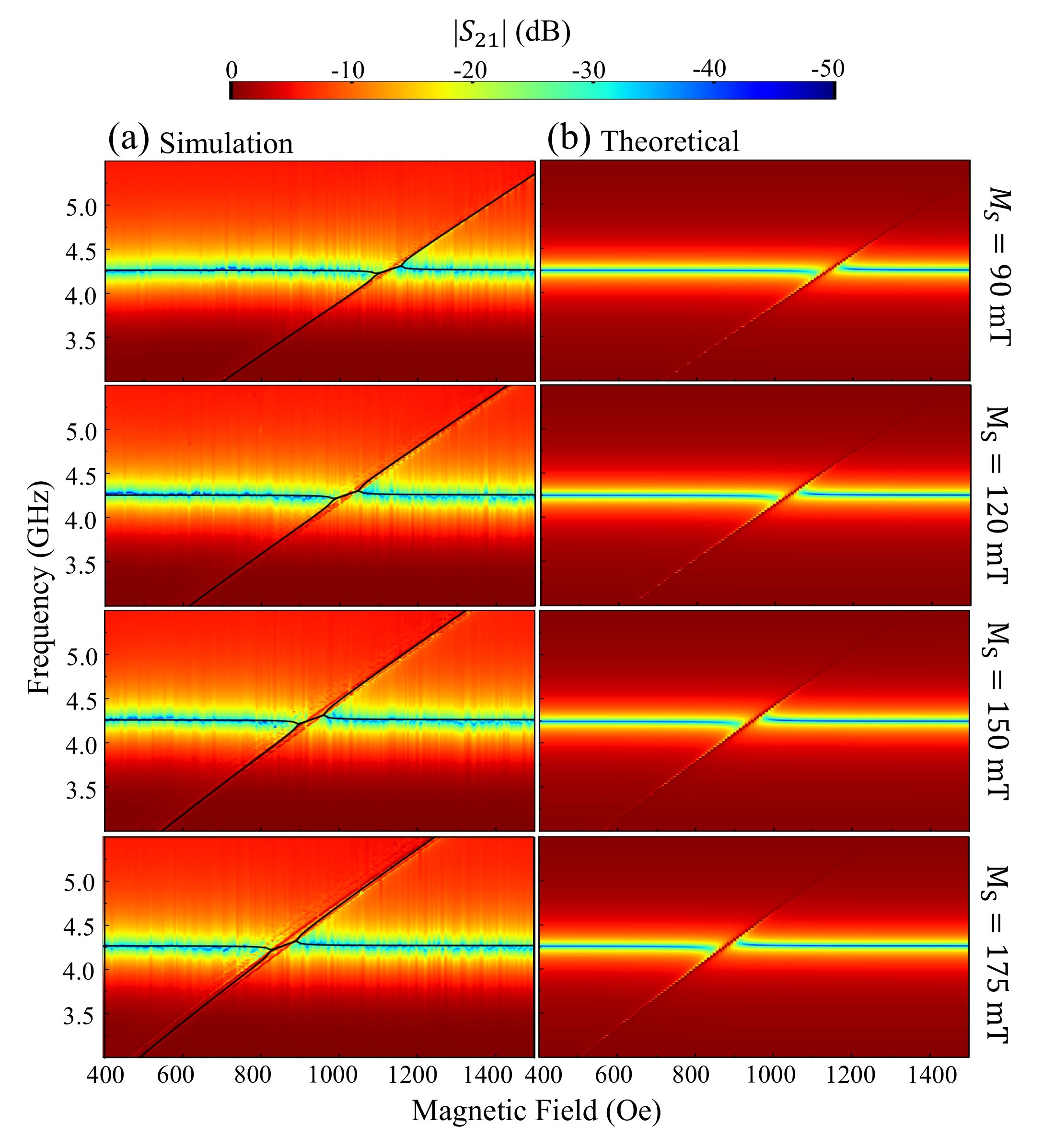}
    
    \caption{ Simulated, fitted, and theoretical $|S_{21}|$ spectra ($0$ to $-50$ dB) in the $f$--$H$ plane at fixed magnon damping. Each row corresponds to a different saturation magnetization value: $M_s = 90$, $120$, $150$, and $175$ mT (top to bottom). (a) The first column shows the full-wave simulated $|S_{21}|$ spectra with eigenvalue fitting results overlaid as black solid curves, and (b) shows the corresponding theoretical $|S_{21}|$ spectra.
}
    \label{fig:4}
\end{figure}

In addition to magnon damping, the saturation magnetization $M_s$ plays an important role in determining the photon–magnon hybridization. It influences both the density of participating spins and the ferromagnetic resonance frequency of the YIG film. To examine this dependence, the transmission spectra are calculated for $M_{s}$ = 90 to 175 mT, while the Gilbert damping parameter is varied over the corresponding range considered in the simulations. The resulting spectra are shown in Fig. \ref{fig:4}\textcolor{blue}{(a)}.

At lower $M_{s}$, the modification of the cavity spectrum due to photon–magnon interaction is relatively weak. With increasing $M_{s}$, the hybridization becomes more pronounced, leading to a stronger modification of the resonant spectra. This behavior is consistent with the collective nature of the photon–magnon interaction, for which the coupling strength scales approximately as $g\propto \sqrt{N}$,\cite{Huebl2013} where N is the number of participating spins. Since the saturation magnetization is proportional to the spin density, $M_{s}\propto Ng_{s}{\mu }_{B}/V$ \cite{Tabuchi2014}for a fixed magnetic volume, the collective coupling is expected to increase approximately as $g\propto \sqrt{M_{s}}$, provided that the mode profile and other geometric parameters remain unchanged.

The variation of $M_{s}$ also modifies the magnon resonance frequency through the Kittel relation. Consequently, changing $M_{s}$ alters both the photon–magnon coupling and the detuning between the two modes. As a result, the magnetic field at which the strongest hybridization occurs shifts systematically with $M_{s}$, as observed in Fig. \ref{fig:4}\textcolor{blue}{(a)}. Thus, $M_{s}$ provides an additional means of controlling the resonance condition and hybrid-mode evolution, complementing the control of dissipation through the Gilbert damping parameter.

\section{THEORETICAL MODEL AND PURCELL CONDITION}

To understand the damping and magnetization-dependent behavior observed in the electromagnetic simulations and to establish the conditions for the onset of Purcell-enhanced photon dissipation, we develop a theoretical model of the open photon–magnon hybrid system. The model is constructed to represent the physical configuration used in the CST simulations, namely, a cavity photon mode associated with the IORR, a uniform magnon mode of the YIG film, and their coupling to the electromagnetic environment provided by the microstrip transmission line. This correspondence allows the simulated spectral evolution to be interpreted in terms of the underlying coherent and dissipative interactions.

\subsection{Non-Hermitian Coupled-Mode Model}

A schematic representation of the theoretical model is shown in Fig.  \ref{fig:2}\textcolor{blue}{(b)}. The cavity photon mode, representing the resonant electromagnetic mode of the IORR, is coupled to the uniform magnon mode of the YIG film. Both modes can lose energy to the environment through their respective dissipation channels, while their common coupling to the electromagnetic reservoir gives rise to an environment-mediated dissipative interaction. The cavity decay rate in the model therefore represents the loss of the IORR photon mode, including its radiative coupling to the microstrip environment, whereas the magnon decay rate is determined primarily by the intrinsic magnetic damping of the YIG film. Thus, the structure of the theoretical model directly reflects the physical loss channels present in the simulated hybrid device.

The photon and magnon modes are described by the annihilation (creation) operators $\hat{a}({\hat{a}}^{\dagger })$ and $\hat{b}({\hat{b}}^{\dagger })$, respectively, with corresponding resonance frequencies ${\omega }_{c}$ and ${\omega }_{m}$. Their intrinsic dissipation is incorporated through the complex resonance frequencies ${\widetilde{\omega }}_{c}={\omega }_{c}-i\kappa_{c}$ and ${\widetilde{\omega }}_{m}={\omega }_{m}-i\kappa_{m}$, where $\kappa_{c}=\beta {\omega }_{c}$ and $\kappa_{m}=\alpha {\omega }_{m}$ denote the cavity-photon and magnon energy-decay rates, respectively. The cavity decay rate accounts for the intrinsic and radiative losses of the IORR mode, while $\kappa_{m}$ is associated with the magnetic dissipation of the YIG film and is related to the Gilbert damping parameter $\alpha$. In the CST simulations, variation of $\alpha$ therefore corresponds to a controlled variation of $\kappa_{m}$ in the theoretical description. Similarly, changes in the saturation magnetization $M_s$ modify the magnon resonance frequency and the photon–magnon interaction strength in both descriptions.

The interaction between the photon and magnon modes can occur through two physically distinct channels. The first is coherent coupling, characterized by the coupling strength ($J$), which describes reversible energy exchange between the cavity and magnon modes through their electromagnetic interaction. The second is dissipative coupling, characterized by the parameter ($\Gamma$), which originates from the correlated decay of the two modes through their common electromagnetic environment. These two contributions can be combined into a generally complex effective coupling parameter ($g=J-i\Gamma$). For the IORR–YIG structure considered here, the localized microwave magnetic field provides the physical mechanism for coherent photon–magnon interaction, while the radiative coupling of the open resonator and magnetic mode to the common transmission-line environment provides the corresponding dissipative pathway. The coupled photon–magnon system can consequently be described by the following effective non-Hermitian Hamiltonian,
      
\begin{equation}
H=\hbar\widetilde{\omega}_c \hat{a}^\dagger \hat{a}
+\hbar\widetilde{\omega}_m \hat{b}^\dagger \hat{b}
+\hbar g\left(\hat{a}^\dagger \hat{b}+\hat{b}^\dagger \hat{a}\right)
\label{eq:1}
\end{equation}
where the non-Hermitian terms account for the intrinsic dissipation of the individual modes as well as the environment-mediated dissipative interaction between them. This framework provides a unified description of coherent hybridization and loss-induced mode evolution in the open cavity-magnonic system. The complex eigenfrequencies of the hybrid system are obtained by diagonalizing the non-Hermitian Hamiltonian\cite{Wang2019}:

\begin{equation}
\widetilde{\omega}_{\pm}
=
\frac{\widetilde{\omega}_c+\widetilde{\omega}_m}{2}
\pm
\frac{1}{2}
\sqrt{
\left(
\widetilde{\omega}_c-\widetilde{\omega}_m
\right)^2
+4g^2
}
\label{eq:2}
\end{equation}

In this complex eigenvalues, the real and imaginary parts describe the resonance frequencies and effective decay rates of the coupled modes, respectively. The evolution of these eigenvalues determines the nature of the photon–magnon interaction\cite{Harder2021}. When coherent coupling dominates ($J>\Gamma$), the hybrid spectrum exhibits level repulsion accompanied by normal-mode splitting \cite{Shen2022}. Conversely, when the dissipative interaction becomes dominant ($\Gamma>J$), the coupling is predominantly imaginary, resulting in level attraction, hybrid-mode coalescence, and linewidth evolution \cite{Wang2020}. This transition from coherent to dissipative hybridization forms the theoretical basis for realizing Purcell-enhanced photon dissipation in the non-Hermitian cavity-magnonic system.

\subsection{Dissipative Photon–Magnon Coupling}

For each combination of $M_s$ and $\alpha$, the CST-simulated $S_{21}$ spectra are fitted using the transmission response obtained from Eq.\ref{eq:2}, as shown by the solid curves in Figs. \ref{fig:3}\textcolor{blue}{(a)} and \ref{fig:4}\textcolor{blue}{(a)}. The isolated IORR exhibits a field-independent photon resonance at ${\omega }_{c}/2\pi $ = 4.25 GHz. The fitting procedure provides the cavity decay rate (${\kappa }_{c}/2\pi $), magnon decay rate (${\kappa }_{m}/2\pi$ ), and the effective photon–magnon coupling parameters $g/2\pi$, which are summarized in Table \ref{tab:1}.

The fitting results reveal that the interaction is predominantly dissipative, with the coherent component ($J$) being much smaller than the dissipative component ($\Gamma$). This is consistent with the open geometry of the IORR and its coupling to the common electromagnetic environment through the microstrip line. Importantly, the intrinsic electromagnetic overlap between the IORR and YIG is not altered by varying ‘$\alpha$’; however, the extracted effective interaction and the resulting hybrid-mode response change substantially with the magnon loss. As $\alpha$ increases, the magnon decay rate increases, leading to stronger linewidth broadening and a reduction in the observable hybridization. At the same time, the lossy magnon mode provides an additional channel through which energy initially stored in the photon mode can be dissipated. Consequently, the effective decay of the photon-like mode increases, providing the numerical signature of reservoir-assisted photon dissipation \cite{Zhang2023}.

\begin{table*}
\centering

\scriptsize
\renewcommand{\arraystretch}{1.3}

\begin{tabular}{|c|c|c|c|c|c|c|c|c|c|}
\hline

Sr. No. &
$\alpha/2\pi$ $(\times10^{-4})$ &
$\beta/2\pi$ $(\times10^{-3})$ &
$k_m/2\pi$ (MHz) &
$k_c/2\pi$ (MHz) &

\begin{tabular}[c]{@{}c@{}}
$M_s=1750$ Oe \\
$g/2\pi$ (MHz)
\end{tabular} &

\begin{tabular}[c]{@{}c@{}}
$M_s=1500$ Oe \\
$g/2\pi$ (MHz)
\end{tabular} &

\begin{tabular}[c]{@{}c@{}}
$M_s=1200$ Oe \\
$g/2\pi$ (MHz)
\end{tabular} &

\begin{tabular}[c]{@{}c@{}}
$M_s=900$ Oe \\
$g/2\pi$ (MHz)
\end{tabular} &

\begin{tabular}[c]{@{}c@{}}
$(k_m-k_c)/2$ \\
$\geq g > k_c$
\end{tabular}

\\
\hline

1 & 1.4 & 0.550588 & 0.595 & 2.34 & 54.8 & 53.2 & 50.5 & 48.6 & NO \\
\hline

2 & 14 & 0.778824 & 5.95 & 3.31 & 53.7 & 52.3 & 49.6 & 47.5 & NO \\
\hline

3 & 70 & 0.912941 & 29.75 & 3.88 & 51.4 & 50.7 & 48.4 & 46.7 & NO \\
\hline

4 & 140 & 0.964706 & 59.50 & 4.10 & 49.3 & 48.5 & 45.8 & 44.2 & NO \\
\hline

5 & 210 & 1.072941 & 89.25 & 4.56 & 48.0 & 47.2 & 45.3 & 43.3 & NO \\
\hline

6 & 280 & 1.303529 & 119.00 & 5.54 & 46.1 & 45.4 & 42.6 & 40.7 & YES \\
\hline

\end{tabular}
\caption{Summary of extracted system parameters for different magnon damping values $\alpha$, including the magnon and photon dissipation rates $\kappa_m/2\pi$ and $\kappa_c/2\pi$ and coupling strength $g/2\pi$ for various saturation magnetization values $M_s$. The final column indicates the Purcell condition ($\kappa_m-\kappa_c)/2\geq g>\kappa_c$ .}
\label{tab:1}
\end{table*}

The dependence on $M_s$ provides an independent control of the photon–magnon interaction. As shown in Table \ref{tab:1}, the extracted effective coupling strength increases systematically with $M_s$. This trend is consistent with the collective nature of the photon–magnon interaction, for which the coupling strength scales approximately as $g\propto \sqrt{N}$, with N denoting the number of participating spins. For a fixed magnetic volume, N is proportional to $M_s$, giving the approximate dependence $g\propto \sqrt{M_{s}}$. The increase in $M_s$ also modifies the magnon resonance frequency and hence the magnetic-field position of the photon–magnon resonance, consistent with the evolution observed in the simulated spectra.

\begin{figure}
    \centering
    \includegraphics[width=\columnwidth]{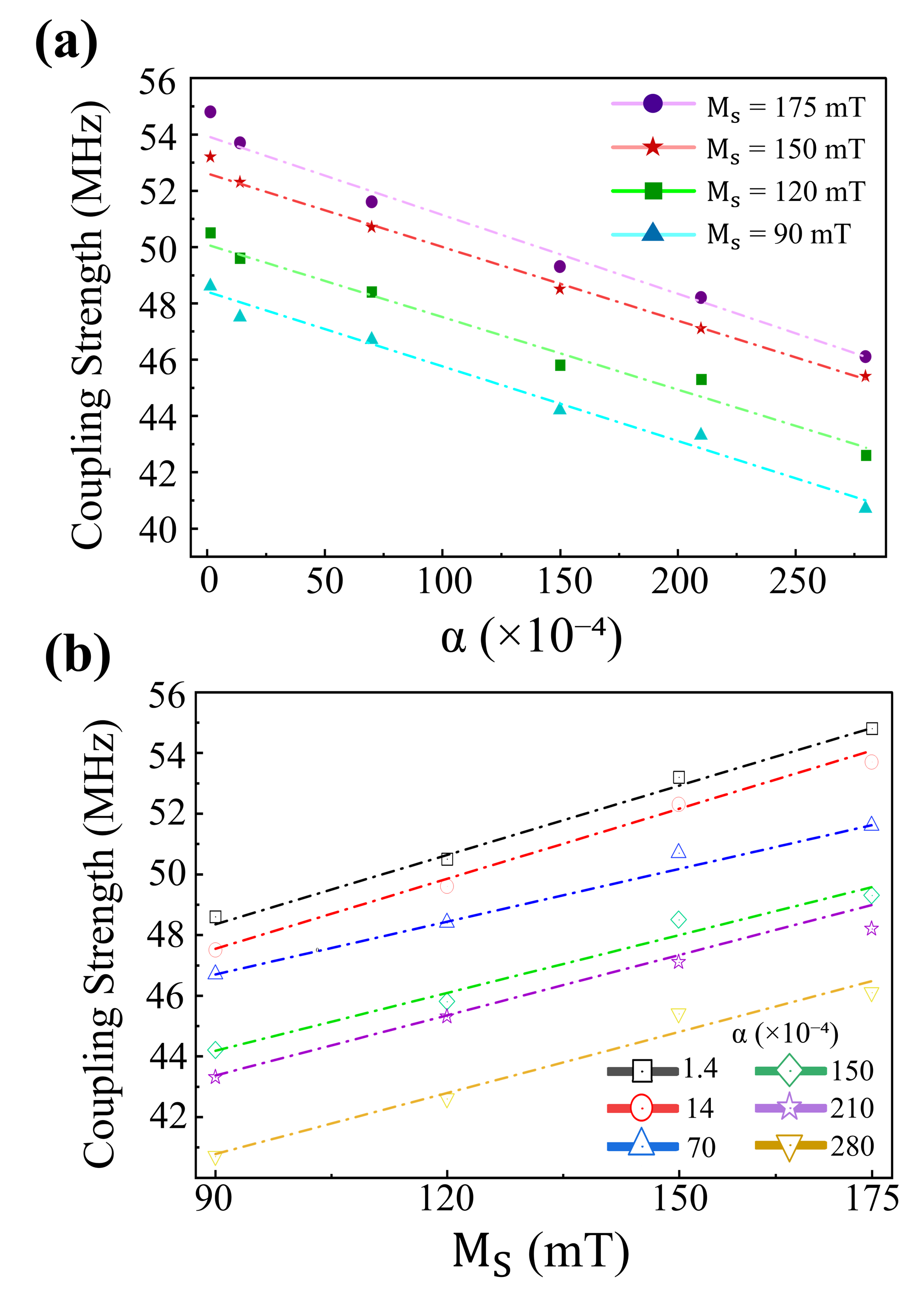}
    \caption{Coupling strength $g/2\pi$ (MHz) as a function of (a)  magnon damping $\alpha$ for different saturation magnetization values and (b) saturation magnetization $M_s$ (mT) for different magnon damping values. }
    \label{fig:5}
\end{figure}

The combined dependence of the extracted interaction parameters on $M_s$ and $\alpha$ is summarized in Figs.\ref{fig:5}\textcolor{blue}{(a)} and \ref{fig:5}\textcolor{blue}{(b)}. The coupling strength increases with increasing $M_s$ and decreases with increasing $\alpha$. These trends demonstrate that the hybrid-mode response can be controlled independently through the magnetic resonance and dissipation of the YIG layer. In the following section, we use these extracted parameters to determine the condition for enhanced photon decay and establish the onset of the Purcell regime within the dissipatively coupled system.

To investigate the driven dynamics of the coupled photon–magnon system, we employ the Heisenberg–Langevin formalism together with the input–output theory. Under the rotating-wave approximation, which is valid because the coupling strength and dissipation rates are much smaller than the resonance frequencies of the photon and magnon modes, the equations of motion are expressed as\cite{Wang2019}

\begin{subequations}\label{eq:3}
\begin{align}
\dot{a} &=
\left[-i\omega_c-(\kappa_c+\kappa)\right]a
-\left(iJ+\Gamma\right)b
+\sqrt{\kappa}\,p_{\mathrm{in}}
\label{eq:3a}
\\
\dot{b} &=
\left[-i\omega_m-(\kappa_m+\gamma)\right]b
-\left(iJ+\Gamma\right)a
+\sqrt{\gamma}\,p_{\mathrm{in}}
\label{eq:3b}
\end{align}
\end{subequations}

where $\kappa$ and $\gamma$ denote the external coupling rates of the cavity and magnon modes to the environment, respectively, and $p_{in}$ represents the incident microwave field. In typical cavity-magnonic experiments, the cavity mode couples much more strongly to the external microwave circuit than the magnon mode, such that $\kappa\gg\gamma$. Under this approximation the excitation of the magnon mode is negligible compared with cavity excitation, allowing the driving term proportional to $\sqrt\gamma$ $p_{in}$ to be is safely neglected. Transforming Eqs. (\ref{eq:3a}) and (\ref{eq:3b}) into the frequency domain yields the following algebraic equations:

\begin{subequations}\label{eq:4}
\begin{align}
\left[
i(\omega-\omega_c)
-(\kappa+\kappa_c)
\right]a
-\left(iJ+\Gamma\right)b
&=
-\sqrt{\kappa}\,p_{\mathrm{in}}
\label{eq:4a}
\\
\left[
i(\omega-\omega_m)
-(\kappa_m+\gamma)
\right]b
-\left(iJ+\Gamma\right)a
&=
0
\label{eq:4b}
\end{align}
\end{subequations}
from which the magnon amplitude can be expressed as

\begin{equation}
b=
\frac{iJ+\Gamma}
{i(\omega-\omega_m)-(\kappa_m+\gamma)}
\,a
\label{eq:5}
\end{equation}

Substituting Eq. (5) into the cavity equation Eq.(\ref{eq:4a}) gives the cavity response as
\begin{equation}
a=
\frac{
-\sqrt{\kappa}\,p_{\mathrm{in}}
}{
i(\omega-\omega_c)
-(\kappa+\kappa_c)
-\dfrac{(iJ+\Gamma)^2}
{i(\omega-\omega_m)-(\kappa_m+\gamma)}
}
\label{eq:6}
\end{equation}

The transmitted microwave field is obtained from the standard input–output relation,

\begin{equation*}
p_{\mathrm{out}}
=
p_{\mathrm{in}}
-
\sqrt{\kappa}\,a
\end{equation*}
from which the forward transmission coefficient is calculated as

\begin{equation*}
S_{21}
=
\frac{p_{\mathrm{out}}}{p_{\mathrm{in}}}
=
1-
\frac{\sqrt{\kappa}\,a}{p_{\mathrm{in}}}
\end{equation*}

Substituting the expression for $a$, the transmission coefficient becomes

\begin{equation}
S_{21}
=
1+
\frac{
\kappa
}{
i(\omega-\omega_c)
-(\kappa+\kappa_c)
+\dfrac{-(iJ+\Gamma)^2}
{i(\omega-\omega_m)-(\kappa_m+\gamma)}
}
\label{eq:7}
\end{equation}

The resulting analytical expression shows that the transmission spectrum is governed by the interplay among photon dissipation, magnon dissipation, and the complex photon–magnon coupling\cite{Wang2019}. 

To further validate the analytical framework, we examine the numerical signatures of the dissipative Photon–Magnon Coupling by evaluating the transmission spectra using Equation \ref{eq:7}. Figure \ref{fig:3}\textcolor{blue}{(b)} and \ref{fig:4}\textcolor{blue}{(b)} shows the calculated $|S_{21}|$ transmission spectra for different values $\alpha$ and $M_{s}$ respectively. This shows the evolution of level attraction accompanied by hybrid-mode coalescence, reflecting the non-Hermitian nature of the interaction. These results closely follow the simulated results in Figure \ref{fig:3}\textcolor{blue}{(a)} and \ref{fig:4}\textcolor{blue}{(a)}, demonstrating a good agreement between the simulations and the theoretical model. The theoretical $|S_{21}|$ spectra provide the physical insight necessary for realizing Purcell-enhanced photon dissipation in the dissipative coupling region of the cavity-magnon platform. These observations demonstrate that from the viewpoint of device engineering, saturation magnetization provides an additional degree of freedom for controlling the operating regime of the hybrid system. Unlike the damping constant, which primarily regulates the dissipation pathways, $M_{s}$ determines the collective interaction strength. Simultaneous optimization of $M_{s}$ and $\alpha$ thus enables independent tuning of the coupling strength, hybrid-mode linewidths, spectral topology, and cavity decay dynamics. This capability is particularly important for designing planar non-Hermitian magnonic devices with controllable Purcell enhancement and tunable dissipation.

\subsection{Time-Domain Dynamics of the Photon–Magnon System}

To further examine the consequences of the enhanced magnon dissipation observed in the frequency-domain spectra, we analyze the temporal decay of the cavity photon population using the parameters extracted from the CST simulations (Table \ref{tab:1}). The dynamics is described within the Markovian open-system framework using a Lindblad master equation \cite{Zhao2025}, which accounts for the dissipation of both the cavity photon and magnon modes. The temporal evolution of the reduced density matrix ($\rho$) is therefore governed by master equation

\begin{equation}
\begin{aligned}
\frac{d\rho}{dt}
&=
\frac{i}{\hbar}[H_{sys},\rho]
+\kappa_c
\left(
2a^\dagger \rho a
-
a^\dagger a \rho
-
\rho a^\dagger a
\right) \\
&\quad
+\kappa_m
\left(
2b^\dagger \rho b
-
b^\dagger b \rho
-
\rho b^\dagger b
\right).
\end{aligned}
\end{equation}

where \begin{equation}
H_{sys} = \hbar\omega_c a^\dagger a
+\hbar\omega_m b^\dagger b
+\hbar g\left(a^\dagger b+b^\dagger a\right)
\end{equation}
From the master equation, we derive the coupled equations governing the time evolution of the expectation values of the system operators:

\begin{align*}
\frac{d\langle a^\dagger a\rangle}{dt}
&=
-ig\langle a^\dagger b\rangle
+ig\langle b^\dagger a\rangle
-2\kappa_c\langle a^\dagger a\rangle
\\
\frac{d\langle b^\dagger b\rangle}{dt}
&=
ig\langle a^\dagger b\rangle
-ig\langle b^\dagger a\rangle
-2\kappa_m\langle b^\dagger b\rangle
\\
\frac{d\langle a^\dagger b\rangle}{dt}
&=
i\omega_c\langle a^\dagger b\rangle
-i\omega_m\langle a^\dagger b\rangle
-ig\langle a^\dagger a\rangle
+ig\langle b^\dagger b\rangle
\\
&\quad
-\kappa_c\langle a^\dagger b\rangle
-\kappa_m\langle a^\dagger b\rangle
\\
\frac{d\langle b^\dagger a\rangle}{dt}
&=
-i\omega_c\langle b^\dagger a\rangle
+i\omega_m\langle b^\dagger a\rangle
+ig\langle a^\dagger a\rangle
-ig\langle b^\dagger b\rangle
\\
&\quad
-\kappa_c\langle b^\dagger a\rangle
-\kappa_m\langle b^\dagger a\rangle
\end{align*}
The coupled dynamical equations for the cavity photon number, magnon number, and correlation terms can be written in matrix form as:
\begin{equation}
\frac{d}{dt}
\begin{pmatrix}
\langle a^\dagger a\rangle \\
\langle b^\dagger b\rangle \\
\langle a^\dagger b\rangle \\
\langle b^\dagger a\rangle
\end{pmatrix}
=
C
\begin{pmatrix}
\langle a^\dagger a\rangle \\
\langle b^\dagger b\rangle \\
\langle a^\dagger b\rangle \\
\langle b^\dagger a\rangle
\end{pmatrix}
\label{10}
\end{equation}
where the coefficient matrix ($C$) is given by
\begin{equation}
C=
\begin{pmatrix}
-2\kappa_c & 0 & -ig & ig \\
0 & -2\kappa_m & ig & -ig \\
-ig & ig & -i\Delta-\kappa & 0 \\
ig & -ig & 0 & i\Delta-\kappa
\end{pmatrix}
\label{11}
\end{equation}
Here, $\Delta=\omega_{m}-\omega_{c}$ represents the detuning between the magnon and cavity resonance frequencies, while
$\kappa=\kappa_c+\kappa_m $ is the total damping rate of the coupled system. The parameter ($g$) denotes the magnon–photon coupling strength. The four eigenvalues of the coefficient matrix are, 
\begin{align*}
\Gamma_1 &= -\kappa-\frac{\delta-\delta'}{2}, \\
\Gamma_2 &= -\kappa+\frac{\delta-\delta'}{2}, \\
\Gamma_3 &= -\kappa-\frac{\delta+\delta'}{2}, \\
\Gamma_4 &= -\kappa+\frac{\delta+\delta'}{2}
\end{align*}
Where $\delta=\sqrt{d^2-4g^2-\Delta^2}$, $\delta^\prime=\sqrt{\delta^2+4\Delta^2d^2}$, and $d=\kappa_m-\kappa_c$.
The coupled equation (\ref{10}) describes the exchange of excitations between the cavity photon and magnon modes together with the dissipation induced by the environment. Solving this equation yields the complete decay dynamics of the hybrid system. For the resonant condition we set  $\Delta$ = 0 and the dynamics of the magnon and cavity photon populations are obtained.

\begin{figure}
    \centering
    \includegraphics[width=\columnwidth]{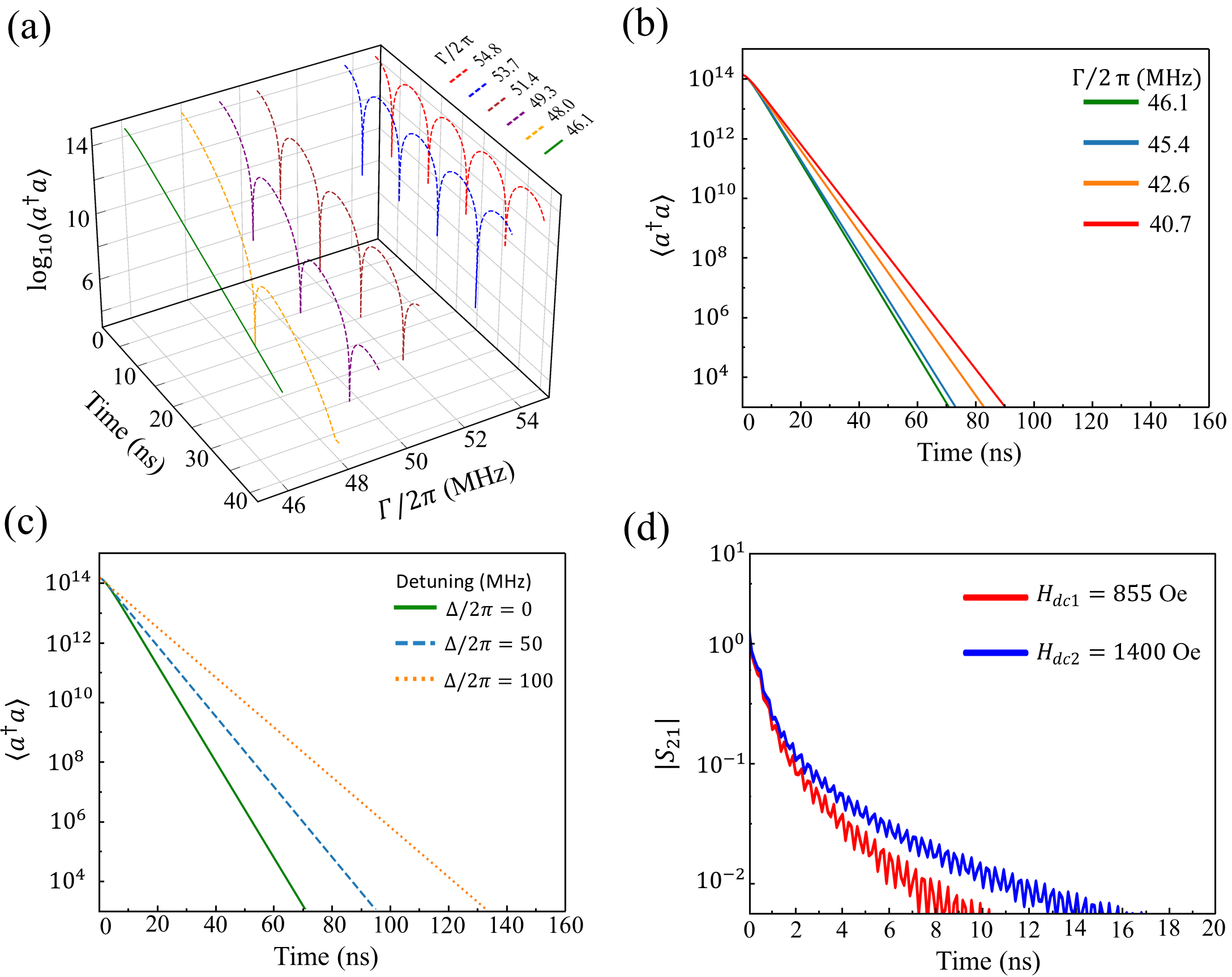}
    \caption{Time evolution of the cavity photon population, $\langle a^\dagger a\rangle$, with an initial cavity photon population $N_c$ and cavity frequency $\omega_c/2\pi = 4.25 $ GHz. (a) Logarithm of the average cavity photon number for different coupling strengths corresponding to the extracted parameters listed in Table I for $M_s$ = 1750 Oe (b) Time evolution of the cavity photon population for different coupling strengths at resonance ($\Delta = 0$) in purcell regime, with $\kappa_c/2\pi = 5.54~\mathrm{MHz}$ and $\kappa_m/2\pi = 119~\mathrm{MHz}$. (c) Time evolution of the cavity photon population for different detuning values, $\Delta/2\pi=0,\ 50,\ 100 $ MHz, with fixed coupling strength $\Gamma/2\pi = 46.1$ MHz and dissipation rates $\kappa_c/2\pi=5.54$ MHz and $\kappa_m/2\pi=119$ MHz. (d) Time-domain evolution of the transmission amplitude $|S_{21}|$ for two different bias magnetic fields, $H_{dc1}=855$\ Oe and $H_{dc2}=1400$\ Oe calculated for $M_s$=1750 Oe at the highest magnon damping $\alpha=2.8\times{10}^{-2}$
}
    \label{fig:6}
\end{figure}

Under the resonant condition $\Delta=0$ (where the photon and magnon modes are maximally interacting), Fig. \ref{fig:6}\textcolor{blue}{(a)} illustrates the time evolution of the cavity photon population $\langle a^{\dagger }a \rangle$ for coupling parameters corresponding to different magnon damping rates. At relatively low magnon damping, the photon population exhibits damped oscillations, reflecting coherent energy exchange between the photon and magnon modes. With increasing magnon damping, these oscillations are progressively suppressed and the decay becomes predominantly monotonic. In the high-damping regime, the magnon mode acts as a strongly lossy channel, causing the excitation transferred from the cavity to decay before significant coherent energy exchange can occur. The transition from oscillatory to monotonic decay therefore provides a clear time-domain signature of the crossover from coherent energy exchange to dissipation-dominated dynamics. A similar trend is obtained for other magnetization values (see Appendix B), where pure decay appears in the highest magnon-damping. When $\delta\geq0$, the system enters the pure decay regime, which corresponds to the condition 
\begin{equation}
g\leq \frac{{\kappa }_{m}-{\kappa }_{c}}{2}
\end{equation}
This regime corresponds to the Purcell regime, where cavity photons decays monotonically without oscillations and this parameter regime coincides with the Purcell condition\cite{Zhao2023APL} introduced in the following section. However, when $\delta\geq0$, then $g > \frac{{\kappa }_{m}-{\kappa }_{c}}{2}$, the system exhibits oscillatory energy exchange between the magnon and cavity photon modes, indicating the transition to the strong-coupling regime. Therefore, the transition from pure decay to oscillatory dynamics serves as a direct signature of the Purcell effect and provides a clear criterion for distinguishing coupling regimes in the cavity–magnon hybrid system.

To further investigate the influence of the coupling strength on the Purcell effect, we study the cavity photon decay dynamics for different coupling strengths within the Purcell regime. The corresponding cavity photon decay dynamics for different coupling strengths are shown in Fig. \ref{fig:6}\textcolor{blue}{(b)}. As the coupling strength decreases, a significant reduction in the cavity decay rate is observed, indicating a suppression of the Purcell effect. Applying a bias magnetic field H shifts the magnon resonance frequency relative to the cavity mode, thereby introducing a finite detuning $\Delta$. Fig. \ref{fig:6}\textcolor{blue}{(c)} shows the cavity photon decay for different detuning values in the Purcell regime. As the detuning increases, the effective interaction between the magnon and cavity modes becomes weaker, leading to suppression of the Purcell effect. Consequently, the overall cavity photon decay rate is reduced, resulting in a slower decay of the cavity population for larger detuning values. The theoretical prediction of enhanced cavity photon dissipation near resonance is further +verified through the time-domain analysis of the simulated $|S_{21}|$ spectra. For the case of $M_{s}$ =1750 Oe and the highest magnon damping $(\alpha =2.8\times {10}^{-2})$, we computed the inverse FFT of the transmission spectra for two different bias magnetic fields, $H_{dc1}$ =855 Oe and $H_{dc2}$ =1400 Oe, and plotted $|S_{21}|$ as a function of time, as shown in Fig. \ref{fig:6}\textcolor{blue}{(d)}. The plot clearly shows that the photon decay rate at the coupling center $(H_{dc1})$ is higher than that of the uncoupled state $(H_{dc2})$. The faster decay observed near the coupling center demonstrates that the cavity photon experiences an additional loss channel when the magnon mode is resonant with the cavity \cite{Zhang2014}. This behavior is consistent with the reservoir-assisted enhancement of photon dissipation predicted by the open-system description.

\subsection{Driven Photon–Magnon Dynamics}

\begin{figure}
    \centering
    \includegraphics[width=\columnwidth]{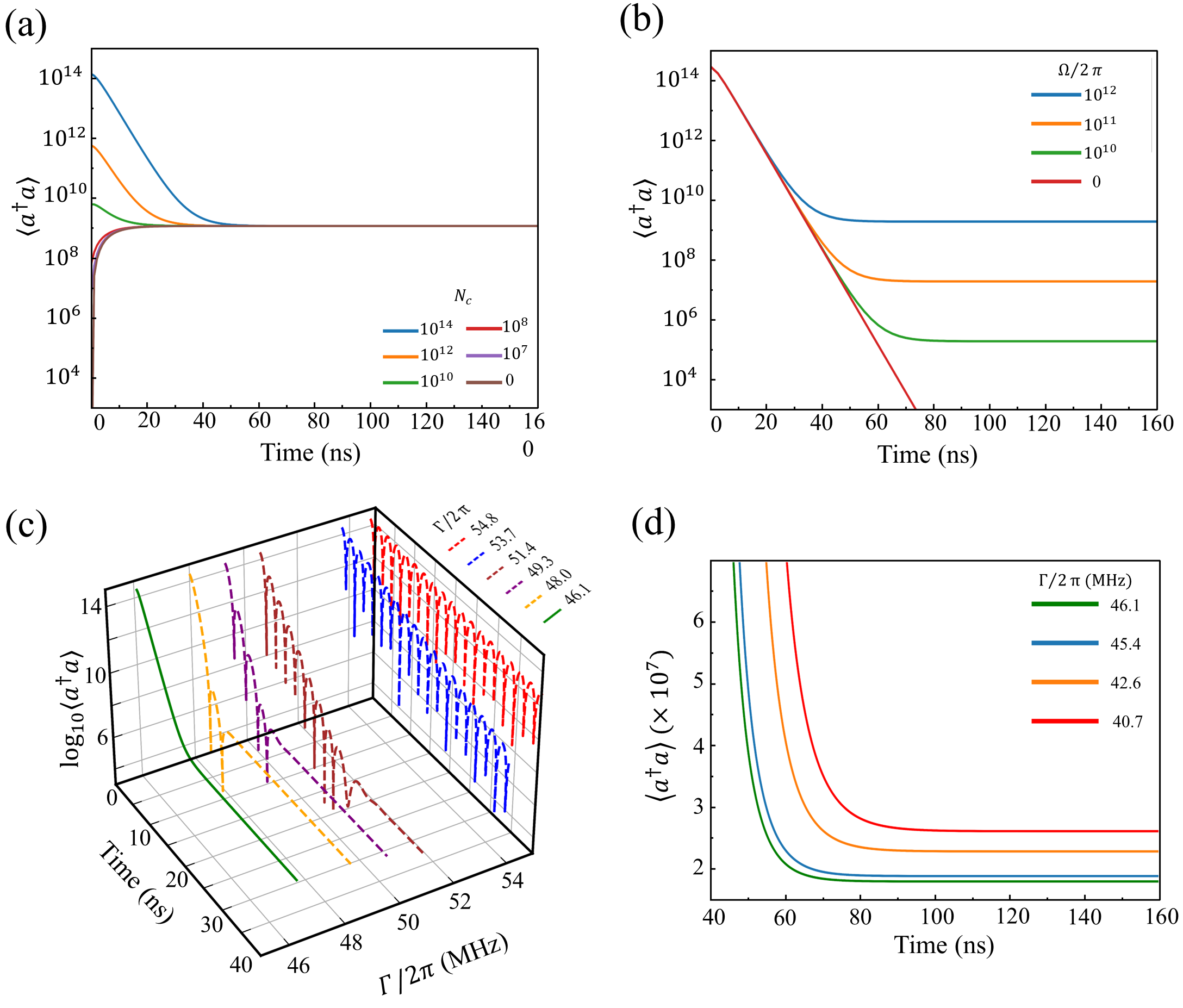}
    \caption{Time evolution of the cavity photon population, $\langle a^\dagger a\rangle$, in the driven cavity–magnon system with cavity frequency $\omega_c/2\pi=4.25$GHz, coupling strength $\Gamma/2\pi=46.1$MHz, and dissipation rates $\kappa_c/2\pi=5.54$MHz and $\kappa_m/2\pi=119$MHz. (a) Cavity photon dynamics for different initial cavity photon populations $N_c$ under coherent driving with fixed driving amplitude $\Omega/2\pi={10}^{12}$Hz. (b) Cavity photon dynamics for different coherent driving amplitudes $\Omega/2\pi={10}^{12},{10}^{11},{10}^{10}$Hz and 0, with fixed initial cavity photon population $N_c$=2.7$\times{10}^{14}$. (c) and (d) are same as Fig. 6(a) and 6(b), respectively, but under coherent driving with driving amplitude $\Omega/2\pi={10}^{11}$Hz.
}
    \label{fig:7}
\end{figure}

The above analysis considers the intrinsic decay of an initially excited hybrid system. To examine whether the dissipation-induced photon decay persists under continuous microwave excitation, we next consider the coherently driven cavity–magnon system. In the presence of an external microwave drive applied to the cavity mode, the system Hamiltonian acquires an additional driving term of the form $i\Omega (a^{\dagger }e^{-i{\omega }_{d}t}-ae^{i{\omega }_{d}t})$, (Appendix A) which continuously drives the hybrid system. The temporal evolution of the cavity photon population is analyzed for different initial cavity photon numbers $N_{c}$  at a fixed driving strength of $\Omega /2\pi ={10}^{12} Hz$ as shown in Fig. \ref{fig:7}\textcolor{blue}{(a)}, the remaining parameters are same as in Fig. \ref{fig:6}\textcolor{blue}{(c)} at resonance. When $N_{c}$ is larger than the steady-state cavity photon number $N_{steady}$ (Eq. (\ref{A4})), the cavity photons exhibit a decay process, whereas for $N_{c} < N_{steady}$, the cavity photon population increases with time. Fig. \ref{fig:7}\textcolor{blue}{(b)} further demonstrates the influence of the driving strength $\Omega$ on the cavity photon decay dynamics for a fixed initial cavity photon population $N_{c} > N_{steady}$. It is observed that reducing the driving strength enhances the decay rate of the cavity photons. In the absence of the external drive $(\Omega=0)$, the dynamics reduce to the non-driven case discussed previously.

Fig. \ref{fig:7}\textcolor{blue}{(c)} illustrates the dependence of the cavity photon decay dynamics on the coupling strength for a fixed driving amplitude $\Omega /2\pi ={10}^{11}$ Hz and the remaining parameters are same as in Fig. \ref{fig:6}\textcolor{blue}{(a)}. Within the Purcell regime, Fig.\ref{fig:7}\textcolor{blue}{(d)} presents the cavity photon decay dynamics for different coupling strengths. A purely decaying behavior is observed, followed by the evolution toward distinct steady-state values. As indicated by Eq.  [\ref{A4}] (Appendix A), the steady-state cavity photon population explicitly depends on the coupling strength. In addition, increasing the coupling strength accelerates the cavity photon decay dynamics, thereby enhancing the Purcell effect, consistent with the results observed in the absence of driving.

\subsection{Purcell-Enhanced Decay under Two-Step Coherent Driving}
To examine Purcell-enhanced photon dissipation under coherent excitation, we consider a two-step driving protocol in which the cavity–magnon system is initially prepared in the vacuum state, with zero photon and magnon populations. The system is first subjected to a strong microwave drive of ${\Omega }_{2}/2\pi ={10}^{13}$ Hz for 400 ns , allowing the photon and magnon populations to build up and reach a steady state. After $t\geq400$ ns the driving strength is reduced to weaker drive ${\Omega }_{2}/2\pi ={10}^{11}$ Hz, and the subsequent evolution of the cavity photon population is monitored. This protocol allows the decay of a finite photon population to be examined in the presence of continuous external driving. 

\begin{figure}
    \centering
    \includegraphics[width=\columnwidth]{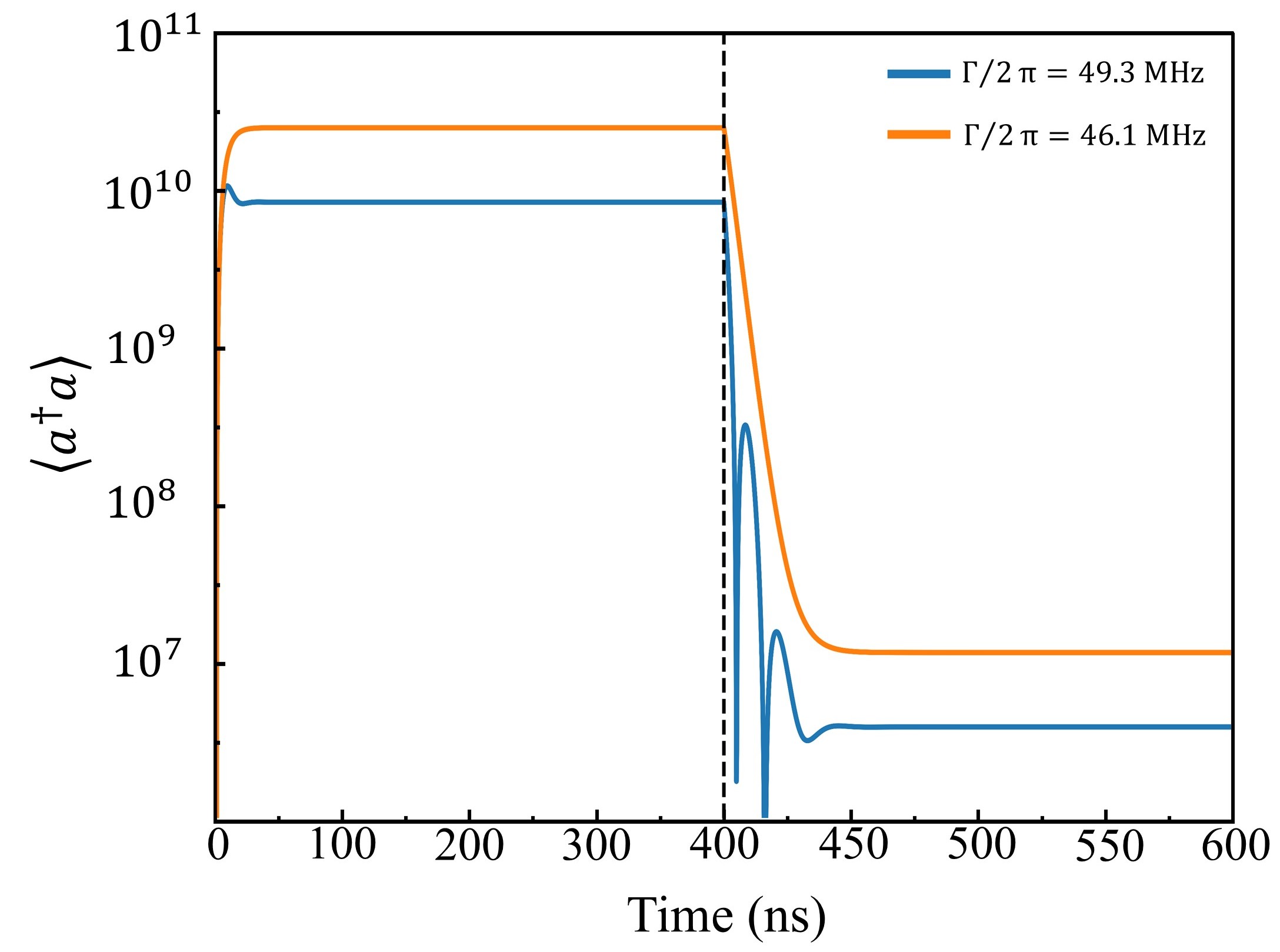}
    \caption{Time evolution of the cavity photon population, $\langle a^\dagger a\rangle$, for two different coupling strengths $\Gamma/2\pi=49.3$MHz and 46.1 MHz under a two-step coherent driving process. The system is initially driven with a strong driving amplitude $\Omega_1/2\pi={10}^{13}$Hz  and at t = 400 ns the driving amplitude is reduced to $\Omega_2/2\pi={10}^{11}$Hz. The corresponding parameters (Table I) are $\left(\kappa_c/2\pi,\kappa_m/2\pi)=(4.10,59.50\right)$MHz and (5.54, 119) MHz for $\Gamma/2\pi=49.3$MHz and 46.1 MHz, respectively.
}
    \label{fig:8}
\end{figure}

Figure \ref{fig:8} shows the resulting cavity photon dynamics for two photon–magnon coupling strengths, $\Gamma/2\pi$ = 46.1 MHz and 49.3 MHz. During the first stage, the cavity population increases from zero and approaches a steady-state value under the strong drive. Following the reduction of the driving strength ( at $t\geq400$ ns), the photon population decreases towards a new steady state. The rate of this decay depends on the photon–magnon coupling strength: stronger coupling produces faster relaxation of the cavity population because it enhances the transfer of photon excitation to the dissipative magnon channel. The observed dependence demonstrates that the additional photon-loss channel associated with the lossy magnon mode persists under coherent driving and is not restricted to the decay of an initially prepared cavity excitation. The two-step protocol therefore provides a complementary time-domain demonstration of reservoir-assisted photon dissipation. Together with the undriven decay dynamics discussed above, these results support the interpretation of the enhanced cavity decay as a consequence of photon-magnon coupling to a dissipative magnonic reservoir. 

\section{DISCUSSIONS}

The numerical results presented above show that increasing magnon damping strongly modifies the hybrid-mode evolution and enhances cavity-photon dissipation. To establish the origin of this behavior and identify the onset of the Purcell regime, we compare the simulated results with the analytical condition derived from the non-Hermitian coupled-mode model. In the present system, the magnon acts as a lossy auxiliary mode coupled to the cavity photon through dissipative photon–magnon interaction. When energy is transferred from the cavity to the magnon and subsequently dissipated before significant coherent back-transfer occurs, the magnon provides an additional decay channel for the cavity. The resulting increase in the effective photon decay rate is the signature of Purcell-enhanced dissipation. 

The derived Purcell condition defines the boundary between coherent and dissipation-dominated dynamics in terms of the cavity damping, magnon damping, and photon–magnon coupling strength. For the dissipative coupling considered here, the Purcell regime is reached when the magnon dissipation is sufficiently large relative to the coupling, while the coupling remains appreciable larger compared with the intrinsic cavity loss, i.e., $({\kappa }_{m}-{\kappa }_{c})/2\geq g> {\kappa }_{c}$\cite{Zhao2023APL}. This criterion describes the regime in which the lossy magnon can efficiently absorb energy from the cavity without sustained coherent energy exchange. We evaluate this condition using the coupling strength ($g$), and damping parameters $({\kappa }_{m})$ and $({\kappa }_{c})$ extracted from the simulated transmission spectra for different saturation magnetizations. The analysis shows that the Purcell regime is reached at sufficiently high magnon damping, where dissipation dominates the hybrid dynamics. The resulting boundary is consistent with the spectral evolution observed in the CST simulations, in which increasing magnon damping progressively suppresses the frequency merging and broadens the hybrid resonances. Thus, the transition from weak coupling to Purcell regime in dissipation-dominated dynamics is governed by the competition between photon–magnon coupling and the relevant loss rates. Notably, the time-domain results provide an independent test of this criterion. At resonance, the cavity-photon population changes from damped oscillatory decay at lower magnon damping to predominantly monotonic decay when the Purcell condition is satisfied. This behavior reflects the suppression of coherent back-and-forth energy exchange as the magnon becomes increasingly dissipative. The agreement between the analytically derived condition and the transition observed in the time-domain dynamics provides direct support for identifying this parameter range as the Purcell regime.

\begin{figure}
    \centering
    \includegraphics[width=\columnwidth]{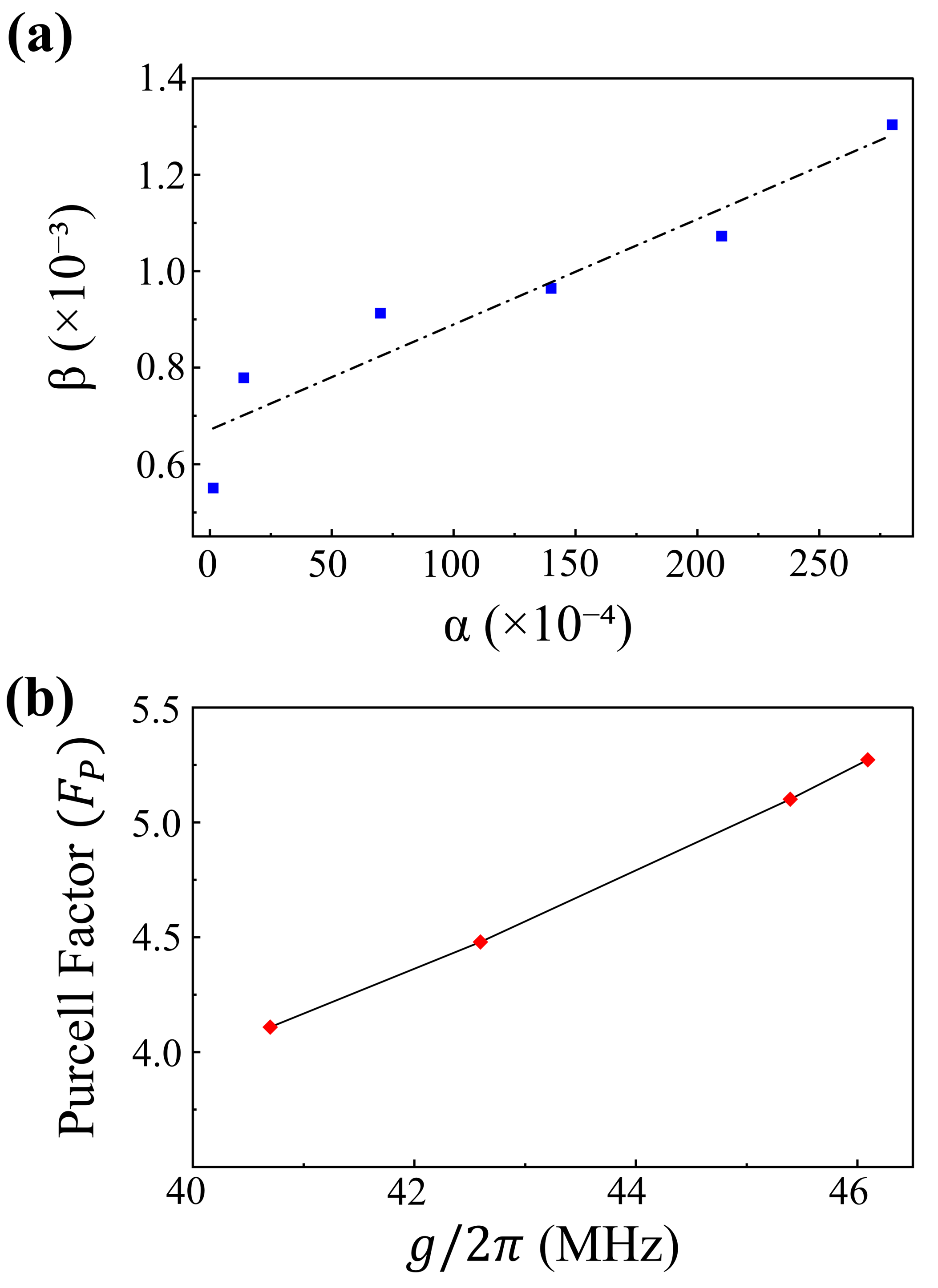}
    \caption{(a) The damping rate of photon (cavity) $\beta$ plotted as a function of the magnon damping rate $\alpha$. (b) Purcell factor $F_P$ as a function of the coupling strength $g/2\pi$, demonstrating the reduction of the Purcell enhancement with decreasing coupling strength }
    \label{fig:9}
\end{figure}

Figure \ref{fig:9}\textcolor{blue}{(a)} shows the extracted effective cavity decay rate ($\beta$) as a function of the intrinsic magnon damping ($\alpha$). The $\beta$ increases systematically with increasing $\alpha$, while the intrinsic cavity loss of the IORR remains essentially unchanged. This behavior indicates that the enhanced photon decay originates from the additional loss channel introduced by the dissipatively coupled magnon mode. As the magnon becomes more strongly damped, energy transferred from the cavity is removed more efficiently, resulting in a shorter effective photon lifetime. The magnitude of the enhancement also depends on the photon–magnon coupling strength. As shown in Fig.  \ref{fig:9}\textcolor{blue}{(b)}, the Purcell factor i.e. $F_{P}=(\kappa -\sqrt{\delta })/2{\kappa }_{c}$\cite{Zhao2023} decreases from approximately 5.27 to 4.11 as the coupling strength decreases from 46.1 to 40.7 MHz. The larger Purcell factor at stronger coupling reflects more efficient transfer of cavity energy into the dissipative magnon channel. For the investigated parameters, the extracted cavity and magnon damping rates are approximately 5.54 MHz and 119 MHz, respectively, and the corresponding coupling strengths satisfy the derived Purcell criterion.
The Purcell regime is not determined by magnon damping alone. The saturation magnetization $M_s$ provides an additional control parameter because it affects both the magnon resonance frequency and the collective photon–magnon coupling strength. As discussed in Sec. 2.3, increasing $M_s$ enhances the coupling, whereas reducing  $M_s$ weakens the interaction. Consequently, changing  $M_s$ can move the system towards or away from the Purcell boundary. For example, at fixed damping (say $\alpha /2\pi =2.10\times {10}^{-2}$), reducing  $M_s$ from 1750 to 900 Oe decreases the coupling strength to ~ 43.3 MHz, approaching the critical value of $({\kappa }_{m}-{\kappa }_{c})/2=42.345$ MHz. The corresponding time-domain response develops weak residual oscillations (Appendix B), indicating proximity to the boundary between the weak coupling and Purcell regimes. Thus, both magnon damping and saturation magnetization provide effective control parameters for tuning the photon-decay dynamics.

The parameters governing the Purcell regime are experimentally accessible in magnetic thin films\cite{Sharma2018}. The $M_s$ of YIG can be modified through temperature\cite{WillCole2023, Gurjar2021}, stoichiometry, film quality, and controlled chemical substitution \cite{WillCole2023,Gurjar2021,Raad2020,Karami2012,Costa2026}, while the Gilbert damping can be engineered through interface modification, impurity incorporation, film thickness, and fabrication conditions \cite{Trempler2020, Kumar2022, Avdizhiyan2025}. Non-magnetic substitution, such as Ga or Al incorporation, can reduce the spin density and hence $M_s$ \cite{Scheffler2023, Sharma2018b, Okada1991}, whereas rare-earth (e.g., Tb, Dy, Bi, or Er) incorporation can modify the magnetic damping through enhanced relaxation and scattering processes\cite{Okada1991, Sun2013}. Growth techniques such as pulsed laser deposition and liquid-phase epitaxy also provide control over crystallinity and defect concentration, enabling systematic modification of the magnetic parameters \cite{Bhoi2018, Zhang2017, Bhoi2013}. In addition, metallic interfaces and spin-current injection provide further approaches for engineering magnetic damping \cite{Sun2013, Jermain2017, Hyde2016, Kosen2019, Haidar2016}. These established methods indicate that the parameter space explored here is experimentally accessible.

The feasibility of controlling the Purcell regime through magnon dissipation is supported by previous cavity-magnonics experiments \cite{Zhao2023APL, Zhao2025}. In three-dimensional cavity–magnon systems, enhanced magnon damping has been used to increase the photon decay rate \cite{Zhang2014}, while more recent hybrid systems have demonstrated electrical control of magnon damping to access the Purcell regime \cite{Zhao2023APL}. These results provide experimental precedent for dissipation engineering and support the feasibility of implementing the proposed mechanism in a planar photon–magnon architecture.

Taken together, the electromagnetic simulations, non-Hermitian analysis, and time-domain calculations provide a consistent description of Purcell-enhanced photon dissipation in the planar dissipatively coupled photon–magnon system. The common electromagnetic environment enables the lossy magnon mode to act as an additional reservoir for the cavity photon, while the magnon damping and saturation magnetization determine the strength and accessibility of this decay channel. The agreement between the derived Purcell criterion, spectral evolution, enhanced effective cavity decay, and suppression of coherent oscillations establishes the Purcell regime in the investigated parameter range. These results demonstrate that photon dissipation can be controlled through magnetic loss and photon–magnon coupling, providing a practical framework for dissipation engineering in planar cavity-magnonic and microwave hybrid systems. 

\section{CONCLUSIONS}

In this work, we investigated Purcell-enhanced photon dissipation in a dissipatively coupled planar photon–magnon system comprising an inverted octagon-ring resonator and a YIG thin film. Full-wave CST simulations demonstrate that increasing magnon damping drives the hybrid system from strong coupling regime to purcell regime in level attraction, linewidth broadening, and enhanced cavity-photon decay. Variation of the saturation magnetization provides an additional means of controlling the photon–magnon coupling and the transition between these regimes. To understand these observations, we developed a non-Hermitian coupled-mode model incorporating dissipative photon–magnon coupling and derived the condition for the Purcell regime. The time-domain analysis confirms this condition through the transition from oscillatory to monotonic photon decay. The extracted parameters from the simulations satisfy the Purcell criterion over a broad range of the investigated magnetic parameters, with Purcell factors ranging from approximately 4.11 to 5.27 for coupling strengths of 40.7 – 46.1 MHz. These results establish that the lossy magnon mode acts as an effective reservoir that provides an additional decay channel for cavity photons. Overall, our study demonstrates a practical route for controlling photon dissipation through magnetic damping and photon–magnon coupling regime in planar cavity-magnonic systems.

\section*{Acknowledgments}
This work was supported by the University Grants Commission (UGC) through a research fellowship and supported by Science and Engineering Research Board (SERB), India (Grant
No. SRG/2023/001355), and the Anusandhan
National Research Foundation (ANRF), India
(Sanction Nos. ANRF/IRG/2025/001896/PS and
ANRF/ARG/2025/006596/PS). Additional support was
received from the Council of Science \& Technology, Uttar
Pradesh (CSTUP) under Project IDs 2470 (Sanction
No. CST/D-1520) and 4482 (Sanction No. CST/D-7/8).
S. Verma acknowledges the Ministry of Education,
Government of India, for the Prime Minister’s Research
Fellowship (PMRF ID-1102628).

\textbf{Statements $\&$ Declarations}

\textbf{Funding:} The authors declare that no funds, grants, or other support were received during the preparation of this manuscript.

\textbf{Competing Interests:} The authors declare that they have no competing interests. 

\textbf{Author Contributions:} All authors contributed to the study conception and design. B. B led the work and wrote the manuscript with S. V. The other co-authors read, commented and approved the final manuscript.  

\textbf{Data Availability:} The data that support the findings of this study are available within the article.

\appendix

\section{Driven Cavity–Magnon Dynamics}
To study the Purcell-enhanced decay dynamics under external excitation, we consider the driven cavity–magnon system. In the presence of a coherent microwave drive, the Hamiltonian introduced an additional driving term acting on the cavity mode. The driven Hamiltonian is written as
\begin{align}
H
=&\,
\hbar\omega_c a^\dagger a
+\hbar\omega_m b^\dagger b
+\hbar g\left(a^\dagger b+b^\dagger a\right)
\nonumber
\\
&\quad
+i\Omega\left(a^\dagger e^{-i\omega_dt}-ae^{i\omega_dt}\right)
\tag{A1}
\end{align}
where $\Omega$ and $\omega_d$ denote the amplitude and frequency of the driving field, respectively.
Using the master equation approach, the coupled dynamical equations for the relevant expectation values can be expressed in matrix form as
\begin{widetext}
\begin{equation}
\frac{d}{dt}
\begin{pmatrix}
\langle a^\dagger a\rangle \\
\langle b^\dagger b\rangle \\
\langle a^\dagger b\rangle \\
\langle b^\dagger a\rangle \\
\langle a^\dagger\rangle' \\
\langle a\rangle' \\
\langle b^\dagger\rangle' \\
\langle b\rangle'
\end{pmatrix}
=
D
\begin{pmatrix}
\langle a^\dagger a\rangle \\
\langle b^\dagger b\rangle \\
\langle a^\dagger b\rangle \\
\langle b^\dagger a\rangle \\
\langle a^\dagger\rangle' \\
\langle a\rangle' \\
\langle b^\dagger\rangle' \\
\langle b\rangle'
\end{pmatrix}
+
\begin{pmatrix}
0 \\
0 \\
0 \\
0 \\
\Omega \\
\Omega \\
0 \\
0
\end{pmatrix}
\tag{A2}
\end{equation}

where,

\begin{align*}
\langle a^\dagger\rangle'
&=
\langle a^\dagger\rangle e^{-i\omega_dt},
&
\langle a\rangle'
&=
\langle a\rangle e^{i\omega_dt},
\\
\langle b^\dagger\rangle'
&=
\langle b^\dagger\rangle e^{-i\omega_dt},
&
\langle b\rangle'
&=
\langle b\rangle e^{i\omega_dt}.
\end{align*}

The coefficient matrix D is given by

\begin{equation}
\resizebox{\textwidth}{!}{$
D=
\begin{pmatrix}
-2\kappa_c & 0 & -ig & ig & \Omega & \Omega & 0 & 0 \\
0 & -2\kappa_m & ig & -ig & 0 & 0 & 0 & 0 \\
-ig & ig & -i\Delta-\kappa & 0 & 0 & 0 & 0 & \Omega \\
ig & -ig & 0 & i\Delta-\kappa & 0 & 0 & \Omega & 0 \\
0 & 0 & 0 & 0 & i(\omega_d-\omega_c)-\kappa_c & 0 & ig & 0 \\
0 & 0 & 0 & 0 & 0 & -i(\omega_d-\omega_c)-\kappa_c & 0 & -ig \\
0 & 0 & 0 & 0 & ig & 0 & i(\omega_d-\omega_m)-\kappa_m & 0 \\
0 & 0 & 0 & 0 & 0 & -ig & 0 & -i(\omega_d-\omega_m)-\kappa_m
\end{pmatrix}
$}
\tag{A3}
\end{equation}

with

\begin{align*}
\Delta &= \omega_m-\omega_c, \\
\kappa &= \kappa_c+\kappa_m .
\end{align*}

In the long-time limit $(t \rightarrow \infty)$,  the steady-state expectation values become constant and are independent of the initial conditions, yielding

\begin{equation}
\langle a^\dagger a\rangle(\infty)
=
\frac{\Omega^2\kappa_m^2}
{(g^2+\kappa_c\kappa_m)^2}
\tag{A4}
\label{A4}
\end{equation}

\begin{equation*}
\langle b^\dagger b\rangle(\infty)
=
\frac{\Omega^2 g^2}
{(g^2+\kappa_c\kappa_m)^2}
\end{equation*}

\begin{equation*}
\langle a^\dagger b\rangle(\infty)
=
\langle b^\dagger a\rangle(\infty)
=
\frac{ig\kappa_m\Omega^2}
{(g^2+\kappa_c\kappa_m)^2}
\end{equation*}

\begin{equation*}
\langle a^\dagger\rangle'(\infty)
=
\langle b\rangle'(\infty)
=
\frac{\kappa_m\Omega}
{g^2+\kappa_c\kappa_m}
\end{equation*}

\begin{equation*}
\langle b^\dagger\rangle'^{*}(\infty)
=
\langle a\rangle'(\infty)
=
\frac{ig\Omega}
{g^2+\kappa_c\kappa_m}
\end{equation*}
\end{widetext}

\clearpage
\begin{widetext}

\section{Additional Results}
This appendix presents additional simulated transmission spectra, fitted eigenvalue dispersions, and parameter-dependent analyses corresponding to different saturation magnetization values beyond the $M_s$ = 175 mT case presented in the main text.

\begin{figure}[b]
\centering
\includegraphics[width=0.9\textwidth]{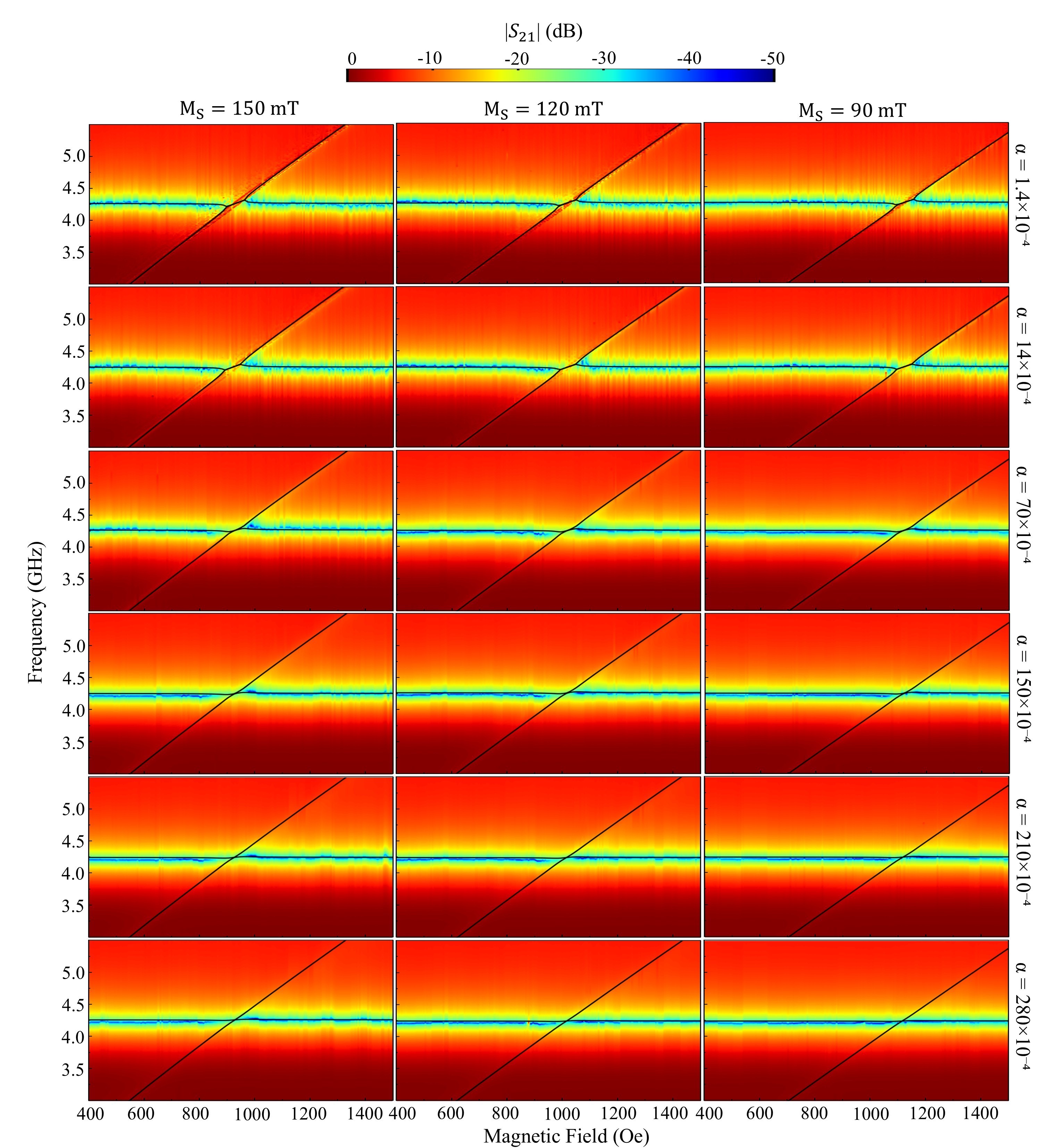}
\caption{
Simulated microwave transmission spectra $\mid S_{21}\mid$ ($0$ to $-50$ dB) in the frequency--magnetic field ($f$--$H$) plane for different magnon damping values $\alpha$ corresponding to $M_s = 1500$, $1200$, and $900$ Oe. The black solid lines represent the fitted eigenvalues.
}
\label{fig10}
\end{figure}

\begin{figure}[b]
\centering
\includegraphics[width=0.9\textwidth]{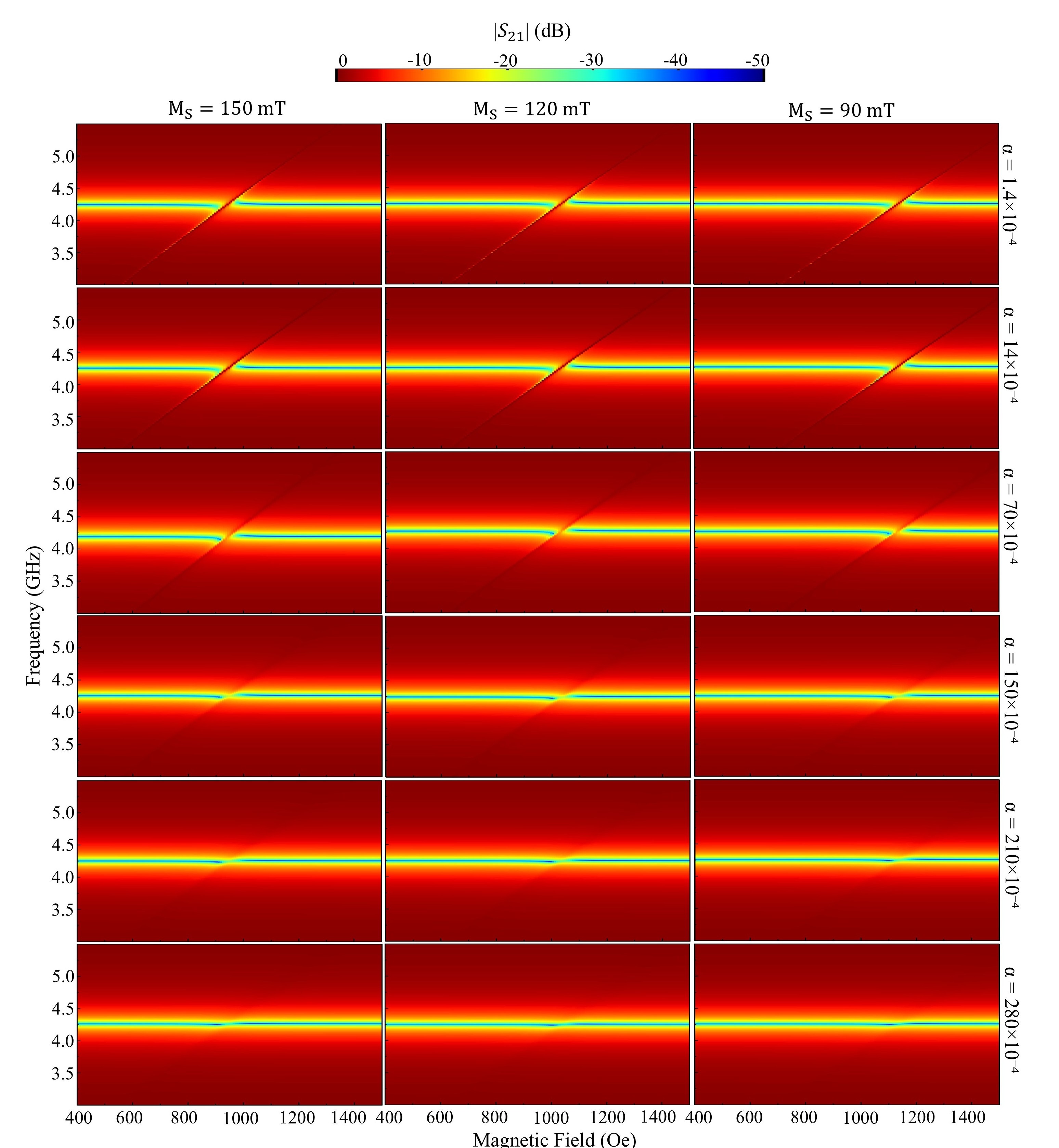}
\caption{ Theoretically fitted microwave transmission spectra $\mid S_{21}\mid$ in the frequency–magnetic field ($f–H$) plane for different magnon damping values $\alpha$ corresponding to $M_s$ = 1500, 1200 and 900 Oe 
}
\label{fig:S1}
\end{figure}

\clearpage
Fig.~\ref{figS3} presents the time evolution of the cavity photon population for different coupling strengths corresponding to (a) $M_s$=1500 Oe, (b) $M_s$=1200 Oe, and (c) $M_s$=900 Oe. For the $M_s$ = 900 Oe case, the coupling strengths $g/2\pi$ = 43.3 MHz and 40.7 MHz lie very close to this transition boundary. Therefore, a magnified comparison is presented in Fig. S3(d). While the $g/2\pi$ = 43.3 MHz case still exhibits oscillatory behavior, the $g/2\pi$ = 40.7 MHz curve shows purely monotonic decay without oscillations, confirming the realization of the Purcell regime.

\begin{figure}
\centering
\includegraphics[width=0.9\textwidth]{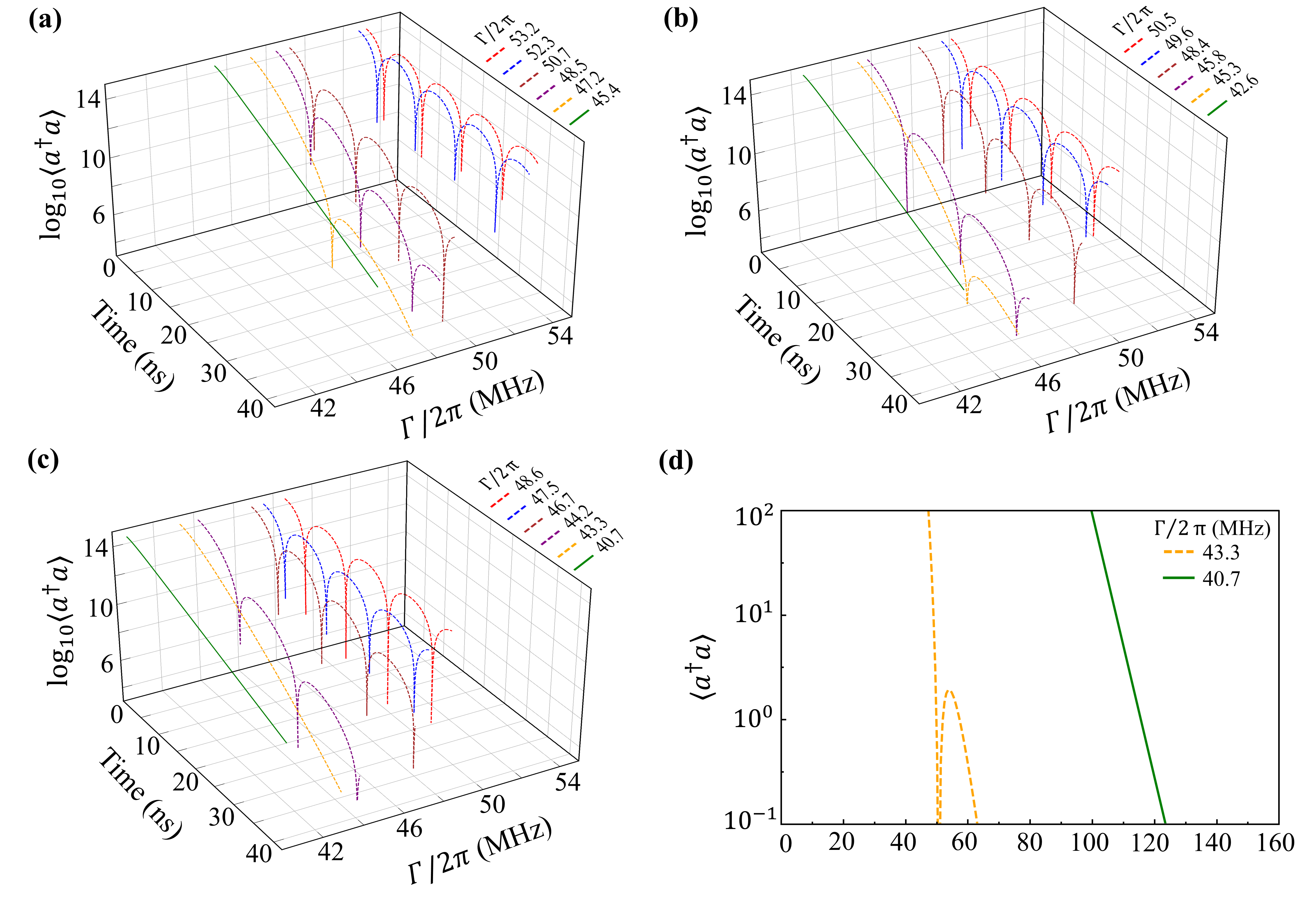}
\caption{
Logarithm of the average cavity photon number for different coupling strengths corresponding to the extracted parameters listed in Table I for (a) $M_s$ = 1500 Oe. (b) $M_s$ = 1200 Oe. (c) $M_s$ = 900 Oe. (d) Time evolution of the cavity photon population at resonance ($\Delta$ = 0) for two different coupling strengths, $g/2\pi$ = 43.3 MHz and 40.7 MHz corresponding to the $M_s$ = 900 Oe}

\label{figS3}
\end{figure}

\end{widetext}
\newpage
\bibliographystyle{apsrev4-2}
\bibliography{reference}

\begin{thebibliography}{68}%
\makeatletter
\providecommand \@ifxundefined [1]{%
 \@ifx{#1\undefined}
}%
\providecommand \@ifnum [1]{%
 \ifnum #1\expandafter \@firstoftwo
 \else \expandafter \@secondoftwo
 \fi
}%
\providecommand \@ifx [1]{%
 \ifx #1\expandafter \@firstoftwo
 \else \expandafter \@secondoftwo
 \fi
}%
\providecommand \natexlab [1]{#1}%
\providecommand \enquote  [1]{``#1''}%
\providecommand \bibnamefont  [1]{#1}%
\providecommand \bibfnamefont [1]{#1}%
\providecommand \citenamefont [1]{#1}%
\providecommand \href@noop [0]{\@secondoftwo}%
\providecommand \href [0]{\begingroup \@sanitize@url \@href}%
\providecommand \@href[1]{\@@startlink{#1}\@@href}%
\providecommand \@@href[1]{\endgroup#1\@@endlink}%
\providecommand \@sanitize@url [0]{\catcode `\\12\catcode `\$12\catcode `\&12\catcode `\#12\catcode `\^12\catcode `\_12\catcode `\%12\relax}%
\providecommand \@@startlink[1]{}%
\providecommand \@@endlink[0]{}%
\providecommand \url  [0]{\begingroup\@sanitize@url \@url }%
\providecommand \@url [1]{\endgroup\@href {#1}{\urlprefix }}%
\providecommand \urlprefix  [0]{URL }%
\providecommand \Eprint [0]{\href }%
\providecommand \doibase [0]{https://doi.org/}%
\providecommand \selectlanguage [0]{\@gobble}%
\providecommand \bibinfo  [0]{\@secondoftwo}%
\providecommand \bibfield  [0]{\@secondoftwo}%
\providecommand \translation [1]{[#1]}%
\providecommand \BibitemOpen [0]{}%
\providecommand \bibitemStop [0]{}%
\providecommand \bibitemNoStop [0]{.\EOS\space}%
\providecommand \EOS [0]{\spacefactor3000\relax}%
\providecommand \BibitemShut  [1]{\csname bibitem#1\endcsname}%
\let\auto@bib@innerbib\@empty
\bibitem [{\citenamefont {Rao}\ \emph {et~al.}(2019)\citenamefont {Rao}, \citenamefont {Yu}, \citenamefont {Zhao}, \citenamefont {Gui}, \citenamefont {Fan}, \citenamefont {Xue},\ and\ \citenamefont {Hu}}]{Rao2019}%
  \BibitemOpen
  \bibfield  {author} {\bibinfo {author} {\bibfnamefont {J.~W.}\ \bibnamefont {Rao}}, \bibinfo {author} {\bibfnamefont {C.~H.}\ \bibnamefont {Yu}}, \bibinfo {author} {\bibfnamefont {Y.~T.}\ \bibnamefont {Zhao}}, \bibinfo {author} {\bibfnamefont {Y.~S.}\ \bibnamefont {Gui}}, \bibinfo {author} {\bibfnamefont {X.~L.}\ \bibnamefont {Fan}}, \bibinfo {author} {\bibfnamefont {D.~S.}\ \bibnamefont {Xue}},\ and\ \bibinfo {author} {\bibfnamefont {C.~M.}\ \bibnamefont {Hu}},\ }\href {https://doi.org/10.1088/1367-2630/ab2482} {\bibfield  {journal} {\bibinfo  {journal} {New Journal of Physics}\ }\textbf {\bibinfo {volume} {21}},\ \bibinfo {pages} {065001} (\bibinfo {year} {2019})}\BibitemShut {NoStop}%
\bibitem [{\citenamefont {Gollwitzer}\ \emph {et~al.}(2021)\citenamefont {Gollwitzer}, \citenamefont {Bocklage}, \citenamefont {Röhlsberger},\ and\ \citenamefont {Meier}}]{Gollwitzer2021}%
  \BibitemOpen
  \bibfield  {author} {\bibinfo {author} {\bibfnamefont {J.}~\bibnamefont {Gollwitzer}}, \bibinfo {author} {\bibfnamefont {L.}~\bibnamefont {Bocklage}}, \bibinfo {author} {\bibfnamefont {R.}~\bibnamefont {Röhlsberger}},\ and\ \bibinfo {author} {\bibfnamefont {G.}~\bibnamefont {Meier}},\ }\href {https://doi.org/10.1038/s41534-021-00445-8} {\bibfield  {journal} {\bibinfo  {journal} {npj Quantum Information}\ }\textbf {\bibinfo {volume} {7}},\ \bibinfo {pages} {114} (\bibinfo {year} {2021})}\BibitemShut {NoStop}%
\bibitem [{\citenamefont {Kurizki}\ \emph {et~al.}(2015)\citenamefont {Kurizki}, \citenamefont {Bertet}, \citenamefont {Kubo}, \citenamefont {Mølmer}, \citenamefont {Petrosyan}, \citenamefont {Rabl},\ and\ \citenamefont {Schmiedmayer}}]{Kurizki2015}%
  \BibitemOpen
  \bibfield  {author} {\bibinfo {author} {\bibfnamefont {G.}~\bibnamefont {Kurizki}}, \bibinfo {author} {\bibfnamefont {P.}~\bibnamefont {Bertet}}, \bibinfo {author} {\bibfnamefont {Y.}~\bibnamefont {Kubo}}, \bibinfo {author} {\bibfnamefont {K.}~\bibnamefont {Mølmer}}, \bibinfo {author} {\bibfnamefont {D.}~\bibnamefont {Petrosyan}}, \bibinfo {author} {\bibfnamefont {P.}~\bibnamefont {Rabl}},\ and\ \bibinfo {author} {\bibfnamefont {J.}~\bibnamefont {Schmiedmayer}},\ }\href {https://doi.org/10.1073/pnas.1419326112} {\bibfield  {journal} {\bibinfo  {journal} {Proceedings of the National Academy of Sciences}\ }\textbf {\bibinfo {volume} {112}},\ \bibinfo {pages} {3866} (\bibinfo {year} {2015})}\BibitemShut {NoStop}%
\bibitem [{\citenamefont {Lachance-Quirion}\ \emph {et~al.}(2019)\citenamefont {Lachance-Quirion}, \citenamefont {Tabuchi}, \citenamefont {Gloppe}, \citenamefont {Usami},\ and\ \citenamefont {Nakamura}}]{LachanceQuirion2019}%
  \BibitemOpen
  \bibfield  {author} {\bibinfo {author} {\bibfnamefont {D.}~\bibnamefont {Lachance-Quirion}}, \bibinfo {author} {\bibfnamefont {Y.}~\bibnamefont {Tabuchi}}, \bibinfo {author} {\bibfnamefont {A.}~\bibnamefont {Gloppe}}, \bibinfo {author} {\bibfnamefont {K.}~\bibnamefont {Usami}},\ and\ \bibinfo {author} {\bibfnamefont {Y.}~\bibnamefont {Nakamura}},\ }\href {https://doi.org/10.7567/1882-0786/ab248d} {\bibfield  {journal} {\bibinfo  {journal} {Applied Physics Express}\ }\textbf {\bibinfo {volume} {12}},\ \bibinfo {pages} {070101} (\bibinfo {year} {2019})}\BibitemShut {NoStop}%
\bibitem [{\citenamefont {Zhang}(2023)}]{Zhang2023}%
  \BibitemOpen
  \bibfield  {author} {\bibinfo {author} {\bibfnamefont {X.}~\bibnamefont {Zhang}},\ }\href {https://doi.org/10.1016/j.mtelec.2023.100044} {\bibfield  {journal} {\bibinfo  {journal} {Materials Today Electronics}\ }\textbf {\bibinfo {volume} {5}},\ \bibinfo {pages} {100044} (\bibinfo {year} {2023})}\BibitemShut {NoStop}%
\bibitem [{\citenamefont {Zare~Rameshti}\ \emph {et~al.}(2022)\citenamefont {Zare~Rameshti}, \citenamefont {Viola~Kusminskiy}, \citenamefont {Haigh}, \citenamefont {Usami}, \citenamefont {Lachance-Quirion}, \citenamefont {Nakamura}, \citenamefont {Hu}, \citenamefont {Tang}, \citenamefont {Bauer},\ and\ \citenamefont {Blanter}}]{ZareRameshti2022}%
  \BibitemOpen
  \bibfield  {author} {\bibinfo {author} {\bibfnamefont {B.}~\bibnamefont {Zare~Rameshti}}, \bibinfo {author} {\bibfnamefont {S.}~\bibnamefont {Viola~Kusminskiy}}, \bibinfo {author} {\bibfnamefont {J.~A.}\ \bibnamefont {Haigh}}, \bibinfo {author} {\bibfnamefont {K.}~\bibnamefont {Usami}}, \bibinfo {author} {\bibfnamefont {D.}~\bibnamefont {Lachance-Quirion}}, \bibinfo {author} {\bibfnamefont {Y.}~\bibnamefont {Nakamura}}, \bibinfo {author} {\bibfnamefont {C.-M.}\ \bibnamefont {Hu}}, \bibinfo {author} {\bibfnamefont {H.~X.}\ \bibnamefont {Tang}}, \bibinfo {author} {\bibfnamefont {G.~E.~W.}\ \bibnamefont {Bauer}},\ and\ \bibinfo {author} {\bibfnamefont {Y.~M.}\ \bibnamefont {Blanter}},\ }\href {https://doi.org/10.1016/j.physrep.2022.06.001} {\bibfield  {journal} {\bibinfo  {journal} {Physics Reports}\ }\textbf {\bibinfo {volume} {979}},\ \bibinfo {pages} {1} (\bibinfo {year} {2022})}\BibitemShut {NoStop}%
\bibitem [{\citenamefont {Boventer}\ \emph {et~al.}(2019)\citenamefont {Boventer}, \citenamefont {Kläui}, \citenamefont {Macêdo},\ and\ \citenamefont {Weides}}]{Boventer2019}%
  \BibitemOpen
  \bibfield  {author} {\bibinfo {author} {\bibfnamefont {I.}~\bibnamefont {Boventer}}, \bibinfo {author} {\bibfnamefont {M.}~\bibnamefont {Kläui}}, \bibinfo {author} {\bibfnamefont {R.}~\bibnamefont {Macêdo}},\ and\ \bibinfo {author} {\bibfnamefont {M.}~\bibnamefont {Weides}},\ }\href {https://doi.org/10.1088/1367-2630/ab5c12} {\bibfield  {journal} {\bibinfo  {journal} {New Journal of Physics}\ }\textbf {\bibinfo {volume} {21}},\ \bibinfo {pages} {125001} (\bibinfo {year} {2019})}\BibitemShut {NoStop}%
\bibitem [{\citenamefont {Junyoung~Kim}(2026)}]{Junyoung2026}%
  \BibitemOpen
  \bibfield  {author} {\bibinfo {author} {\bibfnamefont {S.-K.~K.}\ \bibnamefont {Junyoung~Kim}, \bibfnamefont {Bojong~Kim}},\ }\href {https://doi.org/10.1088/1361-6463/ae74db} {\bibfield  {journal} {\bibinfo  {journal} {Journal of Physics D: Applied Physics}\ }\textbf {\bibinfo {volume} {59}},\ \bibinfo {pages} {193001} (\bibinfo {year} {2026})}\BibitemShut {NoStop}%
\bibitem [{\citenamefont {Xu}\ \emph {et~al.}(2016)\citenamefont {Xu}, \citenamefont {Mason}, \citenamefont {Jiang},\ and\ \citenamefont {Harris}}]{Xu2016}%
  \BibitemOpen
  \bibfield  {author} {\bibinfo {author} {\bibfnamefont {H.}~\bibnamefont {Xu}}, \bibinfo {author} {\bibfnamefont {D.}~\bibnamefont {Mason}}, \bibinfo {author} {\bibfnamefont {L.}~\bibnamefont {Jiang}},\ and\ \bibinfo {author} {\bibfnamefont {J.~G.~E.}\ \bibnamefont {Harris}},\ }\href {https://doi.org/10.1038/nature18604} {\bibfield  {journal} {\bibinfo  {journal} {Nature}\ }\textbf {\bibinfo {volume} {537}},\ \bibinfo {pages} {80} (\bibinfo {year} {2016})}\BibitemShut {NoStop}%
\bibitem [{\citenamefont {Liu}\ and\ \citenamefont {Zhang}(2020)}]{Liu2020}%
  \BibitemOpen
  \bibfield  {author} {\bibinfo {author} {\bibfnamefont {Z.}~\bibnamefont {Liu}}\ and\ \bibinfo {author} {\bibfnamefont {H.}~\bibnamefont {Zhang}},\ }\href {https://doi.org/10.1063/1.5144982} {\bibfield  {journal} {\bibinfo  {journal} {Journal of Applied Physics}\ }\textbf {\bibinfo {volume} {127}},\ \bibinfo {pages} {130901} (\bibinfo {year} {2020})}\BibitemShut {NoStop}%
\bibitem [{\citenamefont {Mi}\ \emph {et~al.}(2025)\citenamefont {Mi}, \citenamefont {Yan}, \citenamefont {Yao}, \citenamefont {Yan}, \citenamefont {Rao},\ and\ \citenamefont {Bai}}]{Mi2025}%
  \BibitemOpen
  \bibfield  {author} {\bibinfo {author} {\bibfnamefont {X.}~\bibnamefont {Mi}}, \bibinfo {author} {\bibfnamefont {L.}~\bibnamefont {Yan}}, \bibinfo {author} {\bibfnamefont {B.}~\bibnamefont {Yao}}, \bibinfo {author} {\bibfnamefont {S.}~\bibnamefont {Yan}}, \bibinfo {author} {\bibfnamefont {J.}~\bibnamefont {Rao}},\ and\ \bibinfo {author} {\bibfnamefont {L.}~\bibnamefont {Bai}},\ }\href {https://doi.org/10.1088/1674-1056/add4e4} {\bibfield  {journal} {\bibinfo  {journal} {Chinese Physics B}\ }\textbf {\bibinfo {volume} {34}},\ \bibinfo {pages} {067508} (\bibinfo {year} {2025})}\BibitemShut {NoStop}%
\bibitem [{\citenamefont {Nair}\ \emph {et~al.}(2022)\citenamefont {Nair}, \citenamefont {Mukhopadhyay},\ and\ \citenamefont {Agarwal}}]{Nair2022}%
  \BibitemOpen
  \bibfield  {author} {\bibinfo {author} {\bibfnamefont {J.~M.~P.}\ \bibnamefont {Nair}}, \bibinfo {author} {\bibfnamefont {D.}~\bibnamefont {Mukhopadhyay}},\ and\ \bibinfo {author} {\bibfnamefont {G.~S.}\ \bibnamefont {Agarwal}},\ }\href {https://doi.org/10.1103/PhysRevB.105.214418} {\bibfield  {journal} {\bibinfo  {journal} {Physical Review B}\ }\textbf {\bibinfo {volume} {105}},\ \bibinfo {pages} {214418} (\bibinfo {year} {2022})}\BibitemShut {NoStop}%
\bibitem [{\citenamefont {Han}\ \emph {et~al.}(2024)\citenamefont {Han}, \citenamefont {Wu}, \citenamefont {Yuan}, \citenamefont {Chen}, \citenamefont {Xia}, \citenamefont {Jiang},\ and\ \citenamefont {Song}}]{Han2024}%
  \BibitemOpen
  \bibfield  {author} {\bibinfo {author} {\bibfnamefont {J.-X.}\ \bibnamefont {Han}}, \bibinfo {author} {\bibfnamefont {J.-L.}\ \bibnamefont {Wu}}, \bibinfo {author} {\bibfnamefont {Z.-H.}\ \bibnamefont {Yuan}}, \bibinfo {author} {\bibfnamefont {Y.-J.}\ \bibnamefont {Chen}}, \bibinfo {author} {\bibfnamefont {Y.}~\bibnamefont {Xia}}, \bibinfo {author} {\bibfnamefont {Y.-Y.}\ \bibnamefont {Jiang}},\ and\ \bibinfo {author} {\bibfnamefont {J.}~\bibnamefont {Song}},\ }\href {https://doi.org/10.1103/PhysRevApplied.21.014057} {\bibfield  {journal} {\bibinfo  {journal} {Physical Review Applied}\ }\textbf {\bibinfo {volume} {21}},\ \bibinfo {pages} {014057} (\bibinfo {year} {2024})}\BibitemShut {NoStop}%
\bibitem [{\citenamefont {Harder}\ \emph {et~al.}(2018)\citenamefont {Harder}, \citenamefont {Yang}, \citenamefont {Yao}, \citenamefont {Yu}, \citenamefont {Rao}, \citenamefont {Gui}, \citenamefont {Stamps},\ and\ \citenamefont {Hu}}]{Harder2018}%
  \BibitemOpen
  \bibfield  {author} {\bibinfo {author} {\bibfnamefont {M.}~\bibnamefont {Harder}}, \bibinfo {author} {\bibfnamefont {Y.}~\bibnamefont {Yang}}, \bibinfo {author} {\bibfnamefont {B.~M.}\ \bibnamefont {Yao}}, \bibinfo {author} {\bibfnamefont {C.~H.}\ \bibnamefont {Yu}}, \bibinfo {author} {\bibfnamefont {J.~W.}\ \bibnamefont {Rao}}, \bibinfo {author} {\bibfnamefont {Y.~S.}\ \bibnamefont {Gui}}, \bibinfo {author} {\bibfnamefont {R.~L.}\ \bibnamefont {Stamps}},\ and\ \bibinfo {author} {\bibfnamefont {C.~M.}\ \bibnamefont {Hu}},\ }\href {https://doi.org/10.1103/PhysRevLett.121.137203} {\bibfield  {journal} {\bibinfo  {journal} {Physical Review Letters}\ }\textbf {\bibinfo {volume} {121}},\ \bibinfo {pages} {137203} (\bibinfo {year} {2018})}\BibitemShut {NoStop}%
\bibitem [{\citenamefont {Moslehi}\ \emph {et~al.}(2024)\citenamefont {Moslehi}, \citenamefont {Baghshahi}, \citenamefont {Faghihi},\ and\ \citenamefont {Mirafzali}}]{Mahboobeh2024}%
  \BibitemOpen
  \bibfield  {author} {\bibinfo {author} {\bibfnamefont {M.}~\bibnamefont {Moslehi}}, \bibinfo {author} {\bibfnamefont {H.~R.}\ \bibnamefont {Baghshahi}}, \bibinfo {author} {\bibfnamefont {M.~J.}\ \bibnamefont {Faghihi}},\ and\ \bibinfo {author} {\bibfnamefont {S.~Y.}\ \bibnamefont {Mirafzali}},\ }\href@noop {} {\bibfield  {journal} {\bibinfo  {journal} {Optics and Laser Technology}\ }\textbf {\bibinfo {volume} {168}},\ \bibinfo {pages} {109920} (\bibinfo {year} {2024})}\BibitemShut {NoStop}%
\bibitem [{\citenamefont {Chen}\ \emph {et~al.}(2023)\citenamefont {Chen}, \citenamefont {Fan}, \citenamefont {Xiong}, \citenamefont {Wang},\ and\ \citenamefont {Ye}}]{Chen2023}%
  \BibitemOpen
  \bibfield  {author} {\bibinfo {author} {\bibfnamefont {J.}~\bibnamefont {Chen}}, \bibinfo {author} {\bibfnamefont {X.-G.}\ \bibnamefont {Fan}}, \bibinfo {author} {\bibfnamefont {W.}~\bibnamefont {Xiong}}, \bibinfo {author} {\bibfnamefont {D.}~\bibnamefont {Wang}},\ and\ \bibinfo {author} {\bibfnamefont {L.}~\bibnamefont {Ye}},\ }\href {https://doi.org/10.1103/PhysRevB.108.024105} {\bibfield  {journal} {\bibinfo  {journal} {Physical Review B}\ }\textbf {\bibinfo {volume} {108}},\ \bibinfo {pages} {024105} (\bibinfo {year} {2023})}\BibitemShut {NoStop}%
\bibitem [{\citenamefont {Aspelmeyer}\ \emph {et~al.}(2014)\citenamefont {Aspelmeyer}, \citenamefont {Kippenberg},\ and\ \citenamefont {Marquardt}}]{Aspelmeyer2014}%
  \BibitemOpen
  \bibfield  {author} {\bibinfo {author} {\bibfnamefont {M.}~\bibnamefont {Aspelmeyer}}, \bibinfo {author} {\bibfnamefont {T.~J.}\ \bibnamefont {Kippenberg}},\ and\ \bibinfo {author} {\bibfnamefont {F.}~\bibnamefont {Marquardt}},\ }\href {https://doi.org/10.1103/RevModPhys.86.1391} {\bibfield  {journal} {\bibinfo  {journal} {Reviews of Modern Physics}\ }\textbf {\bibinfo {volume} {86}},\ \bibinfo {pages} {1391} (\bibinfo {year} {2014})}\BibitemShut {NoStop}%
\bibitem [{\citenamefont {Tabuchi}\ \emph {et~al.}(2016)\citenamefont {Tabuchi}, \citenamefont {Ishino}, \citenamefont {Noguchi}, \citenamefont {Ishikawa}, \citenamefont {Yamazaki}, \citenamefont {Usami},\ and\ \citenamefont {Nakamura}}]{Tabuchi2016}%
  \BibitemOpen
  \bibfield  {author} {\bibinfo {author} {\bibfnamefont {Y.}~\bibnamefont {Tabuchi}}, \bibinfo {author} {\bibfnamefont {S.}~\bibnamefont {Ishino}}, \bibinfo {author} {\bibfnamefont {A.}~\bibnamefont {Noguchi}}, \bibinfo {author} {\bibfnamefont {T.}~\bibnamefont {Ishikawa}}, \bibinfo {author} {\bibfnamefont {R.}~\bibnamefont {Yamazaki}}, \bibinfo {author} {\bibfnamefont {K.}~\bibnamefont {Usami}},\ and\ \bibinfo {author} {\bibfnamefont {Y.}~\bibnamefont {Nakamura}},\ }\href {https://doi.org/10.1016/j.crhy.2016.07.009} {\bibfield  {journal} {\bibinfo  {journal} {Comptes Rendus Physique}\ }\textbf {\bibinfo {volume} {17}},\ \bibinfo {pages} {729} (\bibinfo {year} {2016})}\BibitemShut {NoStop}%
\bibitem [{\citenamefont {Morris}\ \emph {et~al.}(2017)\citenamefont {Morris}, \citenamefont {van Loo}, \citenamefont {Kosen},\ and\ \citenamefont {Karenowska}}]{Morris2017}%
  \BibitemOpen
  \bibfield  {author} {\bibinfo {author} {\bibfnamefont {R.~G.~E.}\ \bibnamefont {Morris}}, \bibinfo {author} {\bibfnamefont {A.~F.}\ \bibnamefont {van Loo}}, \bibinfo {author} {\bibfnamefont {S.}~\bibnamefont {Kosen}},\ and\ \bibinfo {author} {\bibfnamefont {A.~D.}\ \bibnamefont {Karenowska}},\ }\href {https://doi.org/10.1038/s41598-017-11835-4} {\bibfield  {journal} {\bibinfo  {journal} {Scientific Reports}\ }\textbf {\bibinfo {volume} {7}},\ \bibinfo {pages} {11511} (\bibinfo {year} {2017})}\BibitemShut {NoStop}%
\bibitem [{\citenamefont {Atat{\"u}re}\ \emph {et~al.}(2018)\citenamefont {Atat{\"u}re}, \citenamefont {Englund}, \citenamefont {Vamivakas}, \citenamefont {Lee},\ and\ \citenamefont {Wrachtrup}}]{Atature2018}%
  \BibitemOpen
  \bibfield  {author} {\bibinfo {author} {\bibfnamefont {M.}~\bibnamefont {Atat{\"u}re}}, \bibinfo {author} {\bibfnamefont {D.}~\bibnamefont {Englund}}, \bibinfo {author} {\bibfnamefont {N.}~\bibnamefont {Vamivakas}}, \bibinfo {author} {\bibfnamefont {S.-Y.}\ \bibnamefont {Lee}},\ and\ \bibinfo {author} {\bibfnamefont {J.}~\bibnamefont {Wrachtrup}},\ }\href {https://doi.org/10.1038/s41578-018-0008-9} {\bibfield  {journal} {\bibinfo  {journal} {Nature Reviews Materials}\ }\textbf {\bibinfo {volume} {3}},\ \bibinfo {pages} {38} (\bibinfo {year} {2018})}\BibitemShut {NoStop}%
\bibitem [{\citenamefont {Banerjee}\ \emph {et~al.}(2025)\citenamefont {Banerjee}, \citenamefont {Bell}, \citenamefont {Ciccarelli}, \citenamefont {Hesjedal}, \citenamefont {Johnson}, \citenamefont {Kurebayashi} \emph {et~al.}}]{Banerjee2025}%
  \BibitemOpen
  \bibfield  {author} {\bibinfo {author} {\bibfnamefont {N.}~\bibnamefont {Banerjee}}, \bibinfo {author} {\bibfnamefont {C.}~\bibnamefont {Bell}}, \bibinfo {author} {\bibfnamefont {T.}~\bibnamefont {Ciccarelli}}, \bibinfo {author} {\bibfnamefont {T.}~\bibnamefont {Hesjedal}}, \bibinfo {author} {\bibfnamefont {F.}~\bibnamefont {Johnson}}, \bibinfo {author} {\bibfnamefont {H.}~\bibnamefont {Kurebayashi}}, \emph {et~al.},\ }\href {https://doi.org/10.1063/5.0294020} {\bibfield  {journal} {\bibinfo  {journal} {Applied Physics Reviews}\ }\textbf {\bibinfo {volume} {12}},\ \bibinfo {pages} {041328} (\bibinfo {year} {2025})}\BibitemShut {NoStop}%
\bibitem [{\citenamefont {Metelmann}\ and\ \citenamefont {Clerk}(2015)}]{Metelmann2015}%
  \BibitemOpen
  \bibfield  {author} {\bibinfo {author} {\bibfnamefont {A.}~\bibnamefont {Metelmann}}\ and\ \bibinfo {author} {\bibfnamefont {A.~A.}\ \bibnamefont {Clerk}},\ }\href {https://doi.org/10.1103/PhysRevX.5.021025} {\bibfield  {journal} {\bibinfo  {journal} {Physical Review X}\ }\textbf {\bibinfo {volume} {5}},\ \bibinfo {pages} {021025} (\bibinfo {year} {2015})}\BibitemShut {NoStop}%
\bibitem [{\citenamefont {Harder}\ \emph {et~al.}(2021)\citenamefont {Harder}, \citenamefont {Yao}, \citenamefont {Gui},\ and\ \citenamefont {Hu}}]{Harder2021}%
  \BibitemOpen
  \bibfield  {author} {\bibinfo {author} {\bibfnamefont {M.}~\bibnamefont {Harder}}, \bibinfo {author} {\bibfnamefont {B.~M.}\ \bibnamefont {Yao}}, \bibinfo {author} {\bibfnamefont {Y.~S.}\ \bibnamefont {Gui}},\ and\ \bibinfo {author} {\bibfnamefont {C.-M.}\ \bibnamefont {Hu}},\ }\href {https://doi.org/10.1063/5.0046202} {\bibfield  {journal} {\bibinfo  {journal} {Journal of Applied Physics}\ }\textbf {\bibinfo {volume} {129}},\ \bibinfo {pages} {201101} (\bibinfo {year} {2021})}\BibitemShut {NoStop}%
\bibitem [{\citenamefont {Shuai}\ \emph {et~al.}(2025)\citenamefont {Shuai}, \citenamefont {Kim}, \citenamefont {Kim}, \citenamefont {Bhavsar},\ and\ \citenamefont {Kim}}]{Shuai2025}%
  \BibitemOpen
  \bibfield  {author} {\bibinfo {author} {\bibfnamefont {J.}~\bibnamefont {Shuai}}, \bibinfo {author} {\bibfnamefont {B.}~\bibnamefont {Kim}}, \bibinfo {author} {\bibfnamefont {J.}~\bibnamefont {Kim}}, \bibinfo {author} {\bibfnamefont {R.}~\bibnamefont {Bhavsar}},\ and\ \bibinfo {author} {\bibfnamefont {S.-K.}\ \bibnamefont {Kim}},\ }\href {https://doi.org/10.1038/s41598-025-15983-w} {\bibfield  {journal} {\bibinfo  {journal} {Scientific Reports}\ }\textbf {\bibinfo {volume} {15}},\ \bibinfo {pages} {30893} (\bibinfo {year} {2025})}\BibitemShut {NoStop}%
\bibitem [{\citenamefont {Maurya}\ \emph {et~al.}(2024)\citenamefont {Maurya}, \citenamefont {Shrivastava}, \citenamefont {Verma}, \citenamefont {Singh},\ and\ \citenamefont {Bhoi}}]{Maurya2024}%
  \BibitemOpen
  \bibfield  {author} {\bibinfo {author} {\bibfnamefont {A.}~\bibnamefont {Maurya}}, \bibinfo {author} {\bibfnamefont {K.~K.}\ \bibnamefont {Shrivastava}}, \bibinfo {author} {\bibfnamefont {S.}~\bibnamefont {Verma}}, \bibinfo {author} {\bibfnamefont {R.}~\bibnamefont {Singh}},\ and\ \bibinfo {author} {\bibfnamefont {B.}~\bibnamefont {Bhoi}},\ }\href {https://doi.org/10.1016/j.chphi.2024.100669} {\bibfield  {journal} {\bibinfo  {journal} {Chemical Physics Impact}\ }\textbf {\bibinfo {volume} {9}},\ \bibinfo {pages} {100669} (\bibinfo {year} {2024})}\BibitemShut {NoStop}%
\bibitem [{\citenamefont {Yuan}\ \emph {et~al.}(2022)\citenamefont {Yuan}, \citenamefont {Cao}, \citenamefont {Kamra}, \citenamefont {Duine},\ and\ \citenamefont {Yan}}]{Yuan2022}%
  \BibitemOpen
  \bibfield  {author} {\bibinfo {author} {\bibfnamefont {H.~Y.}\ \bibnamefont {Yuan}}, \bibinfo {author} {\bibfnamefont {Y.}~\bibnamefont {Cao}}, \bibinfo {author} {\bibfnamefont {A.}~\bibnamefont {Kamra}}, \bibinfo {author} {\bibfnamefont {R.~A.}\ \bibnamefont {Duine}},\ and\ \bibinfo {author} {\bibfnamefont {P.}~\bibnamefont {Yan}},\ }\href {https://doi.org/10.1016/j.physrep.2022.03.002} {\bibfield  {journal} {\bibinfo  {journal} {Physics Reports}\ }\textbf {\bibinfo {volume} {965}},\ \bibinfo {pages} {1} (\bibinfo {year} {2022})}\BibitemShut {NoStop}%
\bibitem [{\citenamefont {Wang}\ \emph {et~al.}(2022)\citenamefont {Wang}, \citenamefont {Li},\ and\ \citenamefont {Jiang}}]{Wang2022}%
  \BibitemOpen
  \bibfield  {author} {\bibinfo {author} {\bibfnamefont {C.-H.}\ \bibnamefont {Wang}}, \bibinfo {author} {\bibfnamefont {F.}~\bibnamefont {Li}},\ and\ \bibinfo {author} {\bibfnamefont {L.}~\bibnamefont {Jiang}},\ }\href {https://doi.org/10.1038/s41467-022-34373-8} {\bibfield  {journal} {\bibinfo  {journal} {Nature Communications}\ }\textbf {\bibinfo {volume} {13}},\ \bibinfo {pages} {6698} (\bibinfo {year} {2022})}\BibitemShut {NoStop}%
\bibitem [{\citenamefont {Heshami}\ \emph {et~al.}(2016)\citenamefont {Heshami}, \citenamefont {England}, \citenamefont {Humphreys}, \citenamefont {Bustard}, \citenamefont {Acosta}, \citenamefont {Nunn},\ and\ \citenamefont {Sussman}}]{Heshami2016}%
  \BibitemOpen
  \bibfield  {author} {\bibinfo {author} {\bibfnamefont {K.}~\bibnamefont {Heshami}}, \bibinfo {author} {\bibfnamefont {D.~G.}\ \bibnamefont {England}}, \bibinfo {author} {\bibfnamefont {P.~C.}\ \bibnamefont {Humphreys}}, \bibinfo {author} {\bibfnamefont {P.~J.}\ \bibnamefont {Bustard}}, \bibinfo {author} {\bibfnamefont {V.~M.}\ \bibnamefont {Acosta}}, \bibinfo {author} {\bibfnamefont {J.}~\bibnamefont {Nunn}},\ and\ \bibinfo {author} {\bibfnamefont {B.~J.}\ \bibnamefont {Sussman}},\ }\href {https://doi.org/10.1080/09500340.2016.1148212} {\bibfield  {journal} {\bibinfo  {journal} {Journal of Modern Optics}\ }\textbf {\bibinfo {volume} {63}},\ \bibinfo {pages} {2005} (\bibinfo {year} {2016})}\BibitemShut {NoStop}%
\bibitem [{\citenamefont {Zhang}\ \emph {et~al.}(2015)\citenamefont {Zhang}, \citenamefont {Zou}, \citenamefont {Zhu}, \citenamefont {Marquardt}, \citenamefont {Jiang},\ and\ \citenamefont {Tang}}]{Zhang2015}%
  \BibitemOpen
  \bibfield  {author} {\bibinfo {author} {\bibfnamefont {X.}~\bibnamefont {Zhang}}, \bibinfo {author} {\bibfnamefont {C.-L.}\ \bibnamefont {Zou}}, \bibinfo {author} {\bibfnamefont {N.}~\bibnamefont {Zhu}}, \bibinfo {author} {\bibfnamefont {F.}~\bibnamefont {Marquardt}}, \bibinfo {author} {\bibfnamefont {L.}~\bibnamefont {Jiang}},\ and\ \bibinfo {author} {\bibfnamefont {H.~X.}\ \bibnamefont {Tang}},\ }\href {https://doi.org/10.1038/ncomms9914} {\bibfield  {journal} {\bibinfo  {journal} {Nature Communications}\ }\textbf {\bibinfo {volume} {6}},\ \bibinfo {pages} {8914} (\bibinfo {year} {2015})}\BibitemShut {NoStop}%
\bibitem [{\citenamefont {Li}\ \emph {et~al.}(2020)\citenamefont {Li}, \citenamefont {Zhang}, \citenamefont {Tyberkevych}, \citenamefont {Kwok}, \citenamefont {Hoffmann},\ and\ \citenamefont {Novosad}}]{Li2020}%
  \BibitemOpen
  \bibfield  {author} {\bibinfo {author} {\bibfnamefont {Y.}~\bibnamefont {Li}}, \bibinfo {author} {\bibfnamefont {W.}~\bibnamefont {Zhang}}, \bibinfo {author} {\bibfnamefont {V.}~\bibnamefont {Tyberkevych}}, \bibinfo {author} {\bibfnamefont {W.-K.}\ \bibnamefont {Kwok}}, \bibinfo {author} {\bibfnamefont {A.}~\bibnamefont {Hoffmann}},\ and\ \bibinfo {author} {\bibfnamefont {V.}~\bibnamefont {Novosad}},\ }\href {https://doi.org/10.1063/5.0020277} {\bibfield  {journal} {\bibinfo  {journal} {Journal of Applied Physics}\ }\textbf {\bibinfo {volume} {128}},\ \bibinfo {pages} {130902} (\bibinfo {year} {2020})}\BibitemShut {NoStop}%
\bibitem [{\citenamefont {Yang}\ \emph {et~al.}(2024)\citenamefont {Yang}, \citenamefont {Li}, \citenamefont {Wang}, \citenamefont {Zuo}, \citenamefont {Lu}, \citenamefont {Jing},\ and\ \citenamefont {Ren}}]{Yang2024}%
  \BibitemOpen
  \bibfield  {author} {\bibinfo {author} {\bibfnamefont {Z.}~\bibnamefont {Yang}}, \bibinfo {author} {\bibfnamefont {Y.}~\bibnamefont {Li}}, \bibinfo {author} {\bibfnamefont {J.}~\bibnamefont {Wang}}, \bibinfo {author} {\bibfnamefont {Y.}~\bibnamefont {Zuo}}, \bibinfo {author} {\bibfnamefont {T.~X.}\ \bibnamefont {Lu}}, \bibinfo {author} {\bibfnamefont {H.}~\bibnamefont {Jing}},\ and\ \bibinfo {author} {\bibfnamefont {C.}~\bibnamefont {Ren}},\ }\href {https://doi.org/10.1364/OE.528688} {\bibfield  {journal} {\bibinfo  {journal} {Optics Express}\ }\textbf {\bibinfo {volume} {32}},\ \bibinfo {pages} {28293} (\bibinfo {year} {2024})}\BibitemShut {NoStop}%
\bibitem [{\citenamefont {Wang}\ and\ \citenamefont {Hu}(2020)}]{Wang2020}%
  \BibitemOpen
  \bibfield  {author} {\bibinfo {author} {\bibfnamefont {Y.-P.}\ \bibnamefont {Wang}}\ and\ \bibinfo {author} {\bibfnamefont {C.-M.}\ \bibnamefont {Hu}},\ }\href@noop {} {\bibfield  {journal} {\bibinfo  {journal} {Journal of Applied Physics}\ }\textbf {\bibinfo {volume} {127}} (\bibinfo {year} {2020})}\BibitemShut {NoStop}%
\bibitem [{\citenamefont {Bhoi}\ \emph {et~al.}(2019)\citenamefont {Bhoi}, \citenamefont {Kim}, \citenamefont {Jang}, \citenamefont {Kim}, \citenamefont {Yang}, \citenamefont {Cho},\ and\ \citenamefont {Kim}}]{Bhoi2019}%
  \BibitemOpen
  \bibfield  {author} {\bibinfo {author} {\bibfnamefont {B.}~\bibnamefont {Bhoi}}, \bibinfo {author} {\bibfnamefont {B.}~\bibnamefont {Kim}}, \bibinfo {author} {\bibfnamefont {S.-H.}\ \bibnamefont {Jang}}, \bibinfo {author} {\bibfnamefont {J.}~\bibnamefont {Kim}}, \bibinfo {author} {\bibfnamefont {J.}~\bibnamefont {Yang}}, \bibinfo {author} {\bibfnamefont {Y.-J.}\ \bibnamefont {Cho}},\ and\ \bibinfo {author} {\bibfnamefont {S.-K.}\ \bibnamefont {Kim}},\ }\href {https://doi.org/10.1103/PhysRevB.99.134426} {\bibfield  {journal} {\bibinfo  {journal} {Physical Review B}\ }\textbf {\bibinfo {volume} {99}},\ \bibinfo {pages} {134426} (\bibinfo {year} {2019})}\BibitemShut {NoStop}%
\bibitem [{\citenamefont {Stanfield}\ \emph {et~al.}(2023)\citenamefont {Stanfield}, \citenamefont {Powell}, \citenamefont {Horsley}, \citenamefont {Sambles},\ and\ \citenamefont {Hibbins}}]{Stanfield2023}%
  \BibitemOpen
  \bibfield  {author} {\bibinfo {author} {\bibfnamefont {L.~D.}\ \bibnamefont {Stanfield}}, \bibinfo {author} {\bibfnamefont {A.~W.}\ \bibnamefont {Powell}}, \bibinfo {author} {\bibfnamefont {S.~A.~R.}\ \bibnamefont {Horsley}}, \bibinfo {author} {\bibfnamefont {J.~R.}\ \bibnamefont {Sambles}},\ and\ \bibinfo {author} {\bibfnamefont {A.~P.}\ \bibnamefont {Hibbins}},\ }\href {https://doi.org/10.1038/s41598-023-32066-w} {\bibfield  {journal} {\bibinfo  {journal} {Scientific Reports}\ }\textbf {\bibinfo {volume} {13}},\ \bibinfo {pages} {5065} (\bibinfo {year} {2023})}\BibitemShut {NoStop}%
\bibitem [{\citenamefont {Auffèves-Garnier}\ \emph {et~al.}(2007)\citenamefont {Auffèves-Garnier}, \citenamefont {Simon}, \citenamefont {Gérard},\ and\ \citenamefont {Poizat}}]{Auffeves2007}%
  \BibitemOpen
  \bibfield  {author} {\bibinfo {author} {\bibfnamefont {A.}~\bibnamefont {Auffèves-Garnier}}, \bibinfo {author} {\bibfnamefont {C.}~\bibnamefont {Simon}}, \bibinfo {author} {\bibfnamefont {J.-M.}\ \bibnamefont {Gérard}},\ and\ \bibinfo {author} {\bibfnamefont {J.-P.}\ \bibnamefont {Poizat}},\ }\href {https://doi.org/10.1103/PhysRevA.75.053823} {\bibfield  {journal} {\bibinfo  {journal} {Physical Review A}\ }\textbf {\bibinfo {volume} {75}},\ \bibinfo {pages} {053823} (\bibinfo {year} {2007})}\BibitemShut {NoStop}%
\bibitem [{\citenamefont {Krasnok}\ \emph {et~al.}(2015)\citenamefont {Krasnok}, \citenamefont {Slobozhanyuk}, \citenamefont {Simovski}, \citenamefont {Tretyakov}, \citenamefont {Poddubny}, \citenamefont {Miroshnichenko}, \citenamefont {Kivshar},\ and\ \citenamefont {Belov}}]{Krasnok2015}%
  \BibitemOpen
  \bibfield  {author} {\bibinfo {author} {\bibfnamefont {A.~E.}\ \bibnamefont {Krasnok}}, \bibinfo {author} {\bibfnamefont {A.~P.}\ \bibnamefont {Slobozhanyuk}}, \bibinfo {author} {\bibfnamefont {C.~R.}\ \bibnamefont {Simovski}}, \bibinfo {author} {\bibfnamefont {S.~A.}\ \bibnamefont {Tretyakov}}, \bibinfo {author} {\bibfnamefont {A.~N.}\ \bibnamefont {Poddubny}}, \bibinfo {author} {\bibfnamefont {A.~E.}\ \bibnamefont {Miroshnichenko}}, \bibinfo {author} {\bibfnamefont {Y.~S.}\ \bibnamefont {Kivshar}},\ and\ \bibinfo {author} {\bibfnamefont {P.~A.}\ \bibnamefont {Belov}},\ }\href {https://doi.org/10.1038/srep12956} {\bibfield  {journal} {\bibinfo  {journal} {Scientific Reports}\ }\textbf {\bibinfo {volume} {5}},\ \bibinfo {pages} {12956} (\bibinfo {year} {2015})}\BibitemShut {NoStop}%
\bibitem [{\citenamefont {Kaupp}\ \emph {et~al.}(2016)\citenamefont {Kaupp}, \citenamefont {Hümmer}, \citenamefont {Mader}, \citenamefont {Schlederer}, \citenamefont {Benedikter}, \citenamefont {Haeusser}, \citenamefont {Chang}, \citenamefont {Fedder}, \citenamefont {Hänsch},\ and\ \citenamefont {Hunger}}]{Kaupp2016}%
  \BibitemOpen
  \bibfield  {author} {\bibinfo {author} {\bibfnamefont {H.}~\bibnamefont {Kaupp}}, \bibinfo {author} {\bibfnamefont {T.}~\bibnamefont {Hümmer}}, \bibinfo {author} {\bibfnamefont {M.}~\bibnamefont {Mader}}, \bibinfo {author} {\bibfnamefont {B.}~\bibnamefont {Schlederer}}, \bibinfo {author} {\bibfnamefont {J.}~\bibnamefont {Benedikter}}, \bibinfo {author} {\bibfnamefont {P.}~\bibnamefont {Haeusser}}, \bibinfo {author} {\bibfnamefont {H.-C.}\ \bibnamefont {Chang}}, \bibinfo {author} {\bibfnamefont {H.}~\bibnamefont {Fedder}}, \bibinfo {author} {\bibfnamefont {T.~W.}\ \bibnamefont {Hänsch}},\ and\ \bibinfo {author} {\bibfnamefont {D.}~\bibnamefont {Hunger}},\ }\href {https://doi.org/10.1103/PhysRevApplied.6.054010} {\bibfield  {journal} {\bibinfo  {journal} {Physical Review Applied}\ }\textbf {\bibinfo {volume} {6}},\ \bibinfo {pages} {054010} (\bibinfo {year} {2016})}\BibitemShut {NoStop}%
\bibitem [{\citenamefont {Zhao}\ and\ \citenamefont {Qian}(2023)}]{Zhao2023}%
  \BibitemOpen
  \bibfield  {author} {\bibinfo {author} {\bibfnamefont {G.}~\bibnamefont {Zhao}}\ and\ \bibinfo {author} {\bibfnamefont {X.}~\bibnamefont {Qian}},\ }in\ \href@noop {} {\emph {\bibinfo {booktitle} {Frontiers in Optics + Laser Science 2023 (FiO, LS)}}}\ (\bibinfo  {publisher} {Optica Publishing Group},\ \bibinfo {address} {Tacoma, Washington},\ \bibinfo {year} {2023})\BibitemShut {NoStop}%
\bibitem [{\citenamefont {Zhao}\ \emph {et~al.}(2023)\citenamefont {Zhao}, \citenamefont {Wang}, \citenamefont {Han}, \citenamefont {Tian}, \citenamefont {Yan}, \citenamefont {Guo}, \citenamefont {Zhai},\ and\ \citenamefont {Bai}}]{Zhao2023APL}%
  \BibitemOpen
  \bibfield  {author} {\bibinfo {author} {\bibfnamefont {Y.}~\bibnamefont {Zhao}}, \bibinfo {author} {\bibfnamefont {L.}~\bibnamefont {Wang}}, \bibinfo {author} {\bibfnamefont {X.}~\bibnamefont {Han}}, \bibinfo {author} {\bibfnamefont {Y.}~\bibnamefont {Tian}}, \bibinfo {author} {\bibfnamefont {S.}~\bibnamefont {Yan}}, \bibinfo {author} {\bibfnamefont {Q.}~\bibnamefont {Guo}}, \bibinfo {author} {\bibfnamefont {Y.}~\bibnamefont {Zhai}},\ and\ \bibinfo {author} {\bibfnamefont {L.}~\bibnamefont {Bai}},\ }\bibfield  {journal} {\bibinfo  {journal} {Applied Physics Letters}\ }\textbf {\bibinfo {volume} {122}},\ \href {https://doi.org/10.1063/5.0153616} {10.1063/5.0153616} (\bibinfo {year} {2023})\BibitemShut {NoStop}%
\bibitem [{\citenamefont {Verma}\ \emph {et~al.}(2026)\citenamefont {Verma}, \citenamefont {Maurya}, \citenamefont {Khan}, \citenamefont {Shrivastava}, \citenamefont {Singh},\ and\ \citenamefont {Bhoi}}]{Verma2026}%
  \BibitemOpen
  \bibfield  {author} {\bibinfo {author} {\bibfnamefont {S.}~\bibnamefont {Verma}}, \bibinfo {author} {\bibfnamefont {A.}~\bibnamefont {Maurya}}, \bibinfo {author} {\bibfnamefont {F.}~\bibnamefont {Khan}}, \bibinfo {author} {\bibfnamefont {K.~K.}\ \bibnamefont {Shrivastava}}, \bibinfo {author} {\bibfnamefont {R.}~\bibnamefont {Singh}},\ and\ \bibinfo {author} {\bibfnamefont {B.}~\bibnamefont {Bhoi}},\ }\href {https://doi.org/10.1002/qute.202500859} {\bibfield  {journal} {\bibinfo  {journal} {Advanced Quantum Technologies}\ }\textbf {\bibinfo {volume} {9}},\ \bibinfo {pages} {e00859} (\bibinfo {year} {2026})}\BibitemShut {NoStop}%
\bibitem [{\citenamefont {Yu}\ \emph {et~al.}(2019)\citenamefont {Yu}, \citenamefont {Wang}, \citenamefont {Yuan},\ and\ \citenamefont {Xiao}}]{Yu2019}%
  \BibitemOpen
  \bibfield  {author} {\bibinfo {author} {\bibfnamefont {W.}~\bibnamefont {Yu}}, \bibinfo {author} {\bibfnamefont {J.}~\bibnamefont {Wang}}, \bibinfo {author} {\bibfnamefont {H.~Y.}\ \bibnamefont {Yuan}},\ and\ \bibinfo {author} {\bibfnamefont {J.}~\bibnamefont {Xiao}},\ }\href {https://doi.org/10.1103/PhysRevLett.123.227201} {\bibfield  {journal} {\bibinfo  {journal} {Physical Review Letters}\ }\textbf {\bibinfo {volume} {123}},\ \bibinfo {pages} {227201} (\bibinfo {year} {2019})}\BibitemShut {NoStop}%
\bibitem [{\citenamefont {Verma}\ \emph {et~al.}(2025)\citenamefont {Verma}, \citenamefont {Mahalik}, \citenamefont {Maurya}, \citenamefont {Singh},\ and\ \citenamefont {Bhoi}}]{Verma2025b}%
  \BibitemOpen
  \bibfield  {author} {\bibinfo {author} {\bibfnamefont {S.}~\bibnamefont {Verma}}, \bibinfo {author} {\bibfnamefont {J.}~\bibnamefont {Mahalik}}, \bibinfo {author} {\bibfnamefont {A.}~\bibnamefont {Maurya}}, \bibinfo {author} {\bibfnamefont {R.}~\bibnamefont {Singh}},\ and\ \bibinfo {author} {\bibfnamefont {B.}~\bibnamefont {Bhoi}},\ }\href@noop {} {\bibfield  {journal} {\bibinfo  {journal} {New Journal of Physics}\ }\textbf {\bibinfo {volume} {27}},\ \bibinfo {pages} {104506} (\bibinfo {year} {2025})}\BibitemShut {NoStop}%
\bibitem [{\citenamefont {Huebl}\ \emph {et~al.}(2013)\citenamefont {Huebl}, \citenamefont {Zollitsch}, \citenamefont {Lotze}, \citenamefont {Hocke}, \citenamefont {Greifenstein}, \citenamefont {Marx}, \citenamefont {Gross},\ and\ \citenamefont {Goennenwein}}]{Huebl2013}%
  \BibitemOpen
  \bibfield  {author} {\bibinfo {author} {\bibfnamefont {H.}~\bibnamefont {Huebl}}, \bibinfo {author} {\bibfnamefont {C.~W.}\ \bibnamefont {Zollitsch}}, \bibinfo {author} {\bibfnamefont {J.}~\bibnamefont {Lotze}}, \bibinfo {author} {\bibfnamefont {F.}~\bibnamefont {Hocke}}, \bibinfo {author} {\bibfnamefont {M.}~\bibnamefont {Greifenstein}}, \bibinfo {author} {\bibfnamefont {A.}~\bibnamefont {Marx}}, \bibinfo {author} {\bibfnamefont {R.}~\bibnamefont {Gross}},\ and\ \bibinfo {author} {\bibfnamefont {S.~T.~B.}\ \bibnamefont {Goennenwein}},\ }\href {https://doi.org/10.1103/PhysRevLett.111.127003} {\bibfield  {journal} {\bibinfo  {journal} {Physical Review Letters}\ }\textbf {\bibinfo {volume} {111}},\ \bibinfo {pages} {127003} (\bibinfo {year} {2013})}\BibitemShut {NoStop}%
\bibitem [{\citenamefont {Tabuchi}\ \emph {et~al.}(2014)\citenamefont {Tabuchi}, \citenamefont {Ishino}, \citenamefont {Ishikawa}, \citenamefont {Yamazaki}, \citenamefont {Usami},\ and\ \citenamefont {Nakamura}}]{Tabuchi2014}%
  \BibitemOpen
  \bibfield  {author} {\bibinfo {author} {\bibfnamefont {Y.}~\bibnamefont {Tabuchi}}, \bibinfo {author} {\bibfnamefont {S.}~\bibnamefont {Ishino}}, \bibinfo {author} {\bibfnamefont {T.}~\bibnamefont {Ishikawa}}, \bibinfo {author} {\bibfnamefont {R.}~\bibnamefont {Yamazaki}}, \bibinfo {author} {\bibfnamefont {K.}~\bibnamefont {Usami}},\ and\ \bibinfo {author} {\bibfnamefont {Y.}~\bibnamefont {Nakamura}},\ }\href {https://doi.org/10.1103/PhysRevLett.113.083603} {\bibfield  {journal} {\bibinfo  {journal} {Physical Review Letters}\ }\textbf {\bibinfo {volume} {113}},\ \bibinfo {pages} {083603} (\bibinfo {year} {2014})}\BibitemShut {NoStop}%
\bibitem [{\citenamefont {Wang}\ \emph {et~al.}(2019)\citenamefont {Wang}, \citenamefont {Rao}, \citenamefont {Yang}, \citenamefont {Xu}, \citenamefont {Gui}, \citenamefont {Yao}, \citenamefont {You},\ and\ \citenamefont {Hu}}]{Wang2019}%
  \BibitemOpen
  \bibfield  {author} {\bibinfo {author} {\bibfnamefont {Y.-P.}\ \bibnamefont {Wang}}, \bibinfo {author} {\bibfnamefont {J.~W.}\ \bibnamefont {Rao}}, \bibinfo {author} {\bibfnamefont {Y.}~\bibnamefont {Yang}}, \bibinfo {author} {\bibfnamefont {P.-C.}\ \bibnamefont {Xu}}, \bibinfo {author} {\bibfnamefont {Y.~S.}\ \bibnamefont {Gui}}, \bibinfo {author} {\bibfnamefont {B.~M.}\ \bibnamefont {Yao}}, \bibinfo {author} {\bibfnamefont {J.~Q.}\ \bibnamefont {You}},\ and\ \bibinfo {author} {\bibfnamefont {C.~M.}\ \bibnamefont {Hu}},\ }\href {https://doi.org/10.1103/PhysRevLett.123.127202} {\bibfield  {journal} {\bibinfo  {journal} {Physical Review Letters}\ }\textbf {\bibinfo {volume} {123}},\ \bibinfo {pages} {127202} (\bibinfo {year} {2019})}\BibitemShut {NoStop}%
\bibitem [{\citenamefont {Shen}\ \emph {et~al.}(2022)\citenamefont {Shen}, \citenamefont {Xu}, \citenamefont {Zhang}, \citenamefont {Zhang}, \citenamefont {Wang}, \citenamefont {Chai}, \citenamefont {Zou}, \citenamefont {Guo},\ and\ \citenamefont {Dong}}]{Shen2022}%
  \BibitemOpen
  \bibfield  {author} {\bibinfo {author} {\bibfnamefont {Z.}~\bibnamefont {Shen}}, \bibinfo {author} {\bibfnamefont {G.-T.}\ \bibnamefont {Xu}}, \bibinfo {author} {\bibfnamefont {M.}~\bibnamefont {Zhang}}, \bibinfo {author} {\bibfnamefont {Y.-L.}\ \bibnamefont {Zhang}}, \bibinfo {author} {\bibfnamefont {Y.}~\bibnamefont {Wang}}, \bibinfo {author} {\bibfnamefont {C.-Z.}\ \bibnamefont {Chai}}, \bibinfo {author} {\bibfnamefont {C.-L.}\ \bibnamefont {Zou}}, \bibinfo {author} {\bibfnamefont {G.-C.}\ \bibnamefont {Guo}},\ and\ \bibinfo {author} {\bibfnamefont {C.-H.}\ \bibnamefont {Dong}},\ }\href {https://doi.org/10.1103/PhysRevLett.129.243601} {\bibfield  {journal} {\bibinfo  {journal} {Physical Review Letters}\ }\textbf {\bibinfo {volume} {129}},\ \bibinfo {pages} {243601} (\bibinfo {year} {2022})}\BibitemShut {NoStop}%
\bibitem [{\citenamefont {Zhao}\ \emph {et~al.}(2025)\citenamefont {Zhao}, \citenamefont {Wang},\ and\ \citenamefont {Qian}}]{Zhao2025}%
  \BibitemOpen
  \bibfield  {author} {\bibinfo {author} {\bibfnamefont {G.}~\bibnamefont {Zhao}}, \bibinfo {author} {\bibfnamefont {Y.}~\bibnamefont {Wang}},\ and\ \bibinfo {author} {\bibfnamefont {X.-F.}\ \bibnamefont {Qian}},\ }\href {https://doi.org/10.1103/9jw6-w9lw} {\bibfield  {journal} {\bibinfo  {journal} {Physical Review B}\ }\textbf {\bibinfo {volume} {111}},\ \bibinfo {pages} {214428} (\bibinfo {year} {2025})}\BibitemShut {NoStop}%
\bibitem [{\citenamefont {Zhang}\ \emph {et~al.}(2014)\citenamefont {Zhang}, \citenamefont {Zou}, \citenamefont {Jiang},\ and\ \citenamefont {Tang}}]{Zhang2014}%
  \BibitemOpen
  \bibfield  {author} {\bibinfo {author} {\bibfnamefont {X.}~\bibnamefont {Zhang}}, \bibinfo {author} {\bibfnamefont {C.-L.}\ \bibnamefont {Zou}}, \bibinfo {author} {\bibfnamefont {L.}~\bibnamefont {Jiang}},\ and\ \bibinfo {author} {\bibfnamefont {H.~X.}\ \bibnamefont {Tang}},\ }\href {https://doi.org/10.1103/PhysRevLett.113.156401} {\bibfield  {journal} {\bibinfo  {journal} {Physical Review Letters}\ }\textbf {\bibinfo {volume} {113}},\ \bibinfo {pages} {156401} (\bibinfo {year} {2014})}\BibitemShut {NoStop}%
\bibitem [{\citenamefont {Sharma}\ and\ \citenamefont {Kuanr}(2018{\natexlab{a}})}]{Sharma2018}%
  \BibitemOpen
  \bibfield  {author} {\bibinfo {author} {\bibfnamefont {V.}~\bibnamefont {Sharma}}\ and\ \bibinfo {author} {\bibfnamefont {B.~K.}\ \bibnamefont {Kuanr}},\ }\href {https://doi.org/10.1016/j.jallcom.2018.03.086} {\bibfield  {journal} {\bibinfo  {journal} {Journal of Alloys and Compounds}\ }\textbf {\bibinfo {volume} {748}},\ \bibinfo {pages} {591} (\bibinfo {year} {2018}{\natexlab{a}})}\BibitemShut {NoStop}%
\bibitem [{\citenamefont {Will-Cole}\ \emph {et~al.}(2023)\citenamefont {Will-Cole}, \citenamefont {Hart}, \citenamefont {Lauter}, \citenamefont {Grutter}, \citenamefont {Dubs}, \citenamefont {Lindner}, \citenamefont {Reimann}, \citenamefont {Valdez}, \citenamefont {Pearce}, \citenamefont {Monson}, \citenamefont {Cha}, \citenamefont {Heiman},\ and\ \citenamefont {Sun}}]{WillCole2023}%
  \BibitemOpen
  \bibfield  {author} {\bibinfo {author} {\bibfnamefont {A.~R.}\ \bibnamefont {Will-Cole}}, \bibinfo {author} {\bibfnamefont {J.~L.}\ \bibnamefont {Hart}}, \bibinfo {author} {\bibfnamefont {V.}~\bibnamefont {Lauter}}, \bibinfo {author} {\bibfnamefont {A.}~\bibnamefont {Grutter}}, \bibinfo {author} {\bibfnamefont {C.}~\bibnamefont {Dubs}}, \bibinfo {author} {\bibfnamefont {M.}~\bibnamefont {Lindner}}, \bibinfo {author} {\bibfnamefont {T.}~\bibnamefont {Reimann}}, \bibinfo {author} {\bibfnamefont {N.~R.}\ \bibnamefont {Valdez}}, \bibinfo {author} {\bibfnamefont {C.~J.}\ \bibnamefont {Pearce}}, \bibinfo {author} {\bibfnamefont {T.~C.}\ \bibnamefont {Monson}}, \bibinfo {author} {\bibfnamefont {J.~J.}\ \bibnamefont {Cha}}, \bibinfo {author} {\bibfnamefont {D.}~\bibnamefont {Heiman}},\ and\ \bibinfo {author} {\bibfnamefont {N.~X.}\ \bibnamefont {Sun}},\ }\href {https://doi.org/10.1103/PhysRevMaterials.7.054411} {\bibfield  {journal} {\bibinfo  {journal} {Physical Review Materials}\ }\textbf {\bibinfo {volume}
  {7}},\ \bibinfo {pages} {054411} (\bibinfo {year} {2023})}\BibitemShut {NoStop}%
\bibitem [{\citenamefont {Gurjar}\ \emph {et~al.}(2021)\citenamefont {Gurjar}, \citenamefont {Sharma}, \citenamefont {Patnaik},\ and\ \citenamefont {Kuanr}}]{Gurjar2021}%
  \BibitemOpen
  \bibfield  {author} {\bibinfo {author} {\bibfnamefont {G.}~\bibnamefont {Gurjar}}, \bibinfo {author} {\bibfnamefont {V.}~\bibnamefont {Sharma}}, \bibinfo {author} {\bibfnamefont {S.}~\bibnamefont {Patnaik}},\ and\ \bibinfo {author} {\bibfnamefont {B.~K.}\ \bibnamefont {Kuanr}},\ }\href {https://doi.org/10.1088/2053-1591/ac0311} {\bibfield  {journal} {\bibinfo  {journal} {Materials Research Express}\ }\textbf {\bibinfo {volume} {8}},\ \bibinfo {pages} {066401} (\bibinfo {year} {2021})}\BibitemShut {NoStop}%
\bibitem [{\citenamefont {Raad}\ \emph {et~al.}(2020)\citenamefont {Raad}, \citenamefont {Shokrollahi}, \citenamefont {Basavad},\ and\ \citenamefont {Arab}}]{Raad2020}%
  \BibitemOpen
  \bibfield  {author} {\bibinfo {author} {\bibfnamefont {N.~A.}\ \bibnamefont {Raad}}, \bibinfo {author} {\bibfnamefont {H.}~\bibnamefont {Shokrollahi}}, \bibinfo {author} {\bibfnamefont {M.}~\bibnamefont {Basavad}},\ and\ \bibinfo {author} {\bibfnamefont {S.~M.}\ \bibnamefont {Arab}},\ }\bibfield  {journal} {\bibinfo  {journal} {Ceramics International}\ }\textbf {\bibinfo {volume} {46}},\ \href {https://doi.org/10.1016/j.ceramint.2020.06.???} {10.1016/j.ceramint.2020.06.???} (\bibinfo {year} {2020})\BibitemShut {NoStop}%
\bibitem [{\citenamefont {Karami}\ \emph {et~al.}(2012)\citenamefont {Karami}, \citenamefont {Shokrollahi},\ and\ \citenamefont {Hashemi}}]{Karami2012}%
  \BibitemOpen
  \bibfield  {author} {\bibinfo {author} {\bibfnamefont {M.~A.}\ \bibnamefont {Karami}}, \bibinfo {author} {\bibfnamefont {H.}~\bibnamefont {Shokrollahi}},\ and\ \bibinfo {author} {\bibfnamefont {B.}~\bibnamefont {Hashemi}},\ }\bibfield  {journal} {\bibinfo  {journal} {Journal of Magnetism and Magnetic Materials}\ }\textbf {\bibinfo {volume} {324}},\ \href {https://doi.org/10.1016/j.jmmm.2012.04.???} {10.1016/j.jmmm.2012.04.???} (\bibinfo {year} {2012})\BibitemShut {NoStop}%
\bibitem [{\citenamefont {Costa}\ \emph {et~al.}(2026)\citenamefont {Costa}, \citenamefont {Claessens}, \citenamefont {Talmelli}, \citenamefont {Tierno} \emph {et~al.}}]{Costa2026}%
  \BibitemOpen
  \bibfield  {author} {\bibinfo {author} {\bibfnamefont {J.~D.}\ \bibnamefont {Costa}}, \bibinfo {author} {\bibfnamefont {N.}~\bibnamefont {Claessens}}, \bibinfo {author} {\bibfnamefont {G.}~\bibnamefont {Talmelli}}, \bibinfo {author} {\bibfnamefont {D.}~\bibnamefont {Tierno}}, \emph {et~al.},\ }\href@noop {} {\bibfield  {journal} {\bibinfo  {journal} {ACS Applied Materials \& Interfaces}\ } (\bibinfo {year} {2026})}\BibitemShut {NoStop}%
\bibitem [{\citenamefont {Trempler}\ \emph {et~al.}(2020)\citenamefont {Trempler}, \citenamefont {Dreyer}, \citenamefont {Geyer}, \citenamefont {Hauser}, \citenamefont {Woltersdorf},\ and\ \citenamefont {Schmidt}}]{Trempler2020}%
  \BibitemOpen
  \bibfield  {author} {\bibinfo {author} {\bibfnamefont {P.}~\bibnamefont {Trempler}}, \bibinfo {author} {\bibfnamefont {R.}~\bibnamefont {Dreyer}}, \bibinfo {author} {\bibfnamefont {P.}~\bibnamefont {Geyer}}, \bibinfo {author} {\bibfnamefont {C.}~\bibnamefont {Hauser}}, \bibinfo {author} {\bibfnamefont {G.}~\bibnamefont {Woltersdorf}},\ and\ \bibinfo {author} {\bibfnamefont {G.}~\bibnamefont {Schmidt}},\ }\href {https://doi.org/10.1063/5.0026120} {\bibfield  {journal} {\bibinfo  {journal} {Applied Physics Letters}\ }\textbf {\bibinfo {volume} {117}},\ \bibinfo {pages} {232401} (\bibinfo {year} {2020})}\BibitemShut {NoStop}%
\bibitem [{\citenamefont {Kumar}\ \emph {et~al.}(2022)\citenamefont {Kumar}, \citenamefont {Samantaray}, \citenamefont {Das}, \citenamefont {Lal}, \citenamefont {Samal},\ and\ \citenamefont {Hossain}}]{Kumar2022}%
  \BibitemOpen
  \bibfield  {author} {\bibinfo {author} {\bibfnamefont {R.}~\bibnamefont {Kumar}}, \bibinfo {author} {\bibfnamefont {B.}~\bibnamefont {Samantaray}}, \bibinfo {author} {\bibfnamefont {S.}~\bibnamefont {Das}}, \bibinfo {author} {\bibfnamefont {K.}~\bibnamefont {Lal}}, \bibinfo {author} {\bibfnamefont {D.}~\bibnamefont {Samal}},\ and\ \bibinfo {author} {\bibfnamefont {Z.}~\bibnamefont {Hossain}},\ }\href {https://doi.org/10.1103/PhysRevB.106.054405} {\bibfield  {journal} {\bibinfo  {journal} {Physical Review B}\ }\textbf {\bibinfo {volume} {106}},\ \bibinfo {pages} {054405} (\bibinfo {year} {2022})}\BibitemShut {NoStop}%
\bibitem [{\citenamefont {Avdizhiyan}\ \emph {et~al.}(2025)\citenamefont {Avdizhiyan}, \citenamefont {Kazlou}, \citenamefont {Kaihara}, \citenamefont {Stupakiewicz},\ and\ \citenamefont {Razdolski}}]{Avdizhiyan2025}%
  \BibitemOpen
  \bibfield  {author} {\bibinfo {author} {\bibfnamefont {A.}~\bibnamefont {Avdizhiyan}}, \bibinfo {author} {\bibfnamefont {A.}~\bibnamefont {Kazlou}}, \bibinfo {author} {\bibfnamefont {T.}~\bibnamefont {Kaihara}}, \bibinfo {author} {\bibfnamefont {A.}~\bibnamefont {Stupakiewicz}},\ and\ \bibinfo {author} {\bibfnamefont {I.}~\bibnamefont {Razdolski}},\ }\href {https://doi.org/10.1021/acsphotonics.5c00933} {\bibfield  {journal} {\bibinfo  {journal} {ACS Photonics}\ }\textbf {\bibinfo {volume} {12}},\ \bibinfo {pages} {5024} (\bibinfo {year} {2025})}\BibitemShut {NoStop}%
\bibitem [{\citenamefont {Scheffler}\ \emph {et~al.}(2023)\citenamefont {Scheffler}, \citenamefont {Steuer}, \citenamefont {Zhou}, \citenamefont {Siegl}, \citenamefont {Goennenwein},\ and\ \citenamefont {Lammel}}]{Scheffler2023}%
  \BibitemOpen
  \bibfield  {author} {\bibinfo {author} {\bibfnamefont {D.}~\bibnamefont {Scheffler}}, \bibinfo {author} {\bibfnamefont {O.}~\bibnamefont {Steuer}}, \bibinfo {author} {\bibfnamefont {S.}~\bibnamefont {Zhou}}, \bibinfo {author} {\bibfnamefont {L.}~\bibnamefont {Siegl}}, \bibinfo {author} {\bibfnamefont {S.~T.~B.}\ \bibnamefont {Goennenwein}},\ and\ \bibinfo {author} {\bibfnamefont {M.}~\bibnamefont {Lammel}},\ }\href {https://doi.org/10.1103/PhysRevMaterials.7.094405} {\bibfield  {journal} {\bibinfo  {journal} {Physical Review Materials}\ }\textbf {\bibinfo {volume} {7}},\ \bibinfo {pages} {094405} (\bibinfo {year} {2023})}\BibitemShut {NoStop}%
\bibitem [{\citenamefont {Sharma}\ and\ \citenamefont {Kuanr}(2018{\natexlab{b}})}]{Sharma2018b}%
  \BibitemOpen
  \bibfield  {author} {\bibinfo {author} {\bibfnamefont {V.}~\bibnamefont {Sharma}}\ and\ \bibinfo {author} {\bibfnamefont {B.}~\bibnamefont {Kuanr}},\ }\bibfield  {journal} {\bibinfo  {journal} {Journal of Alloys and Compounds}\ }\textbf {\bibinfo {volume} {748}},\ \href {https://doi.org/10.1016/j.jallcom.2018.03.086} {10.1016/j.jallcom.2018.03.086} (\bibinfo {year} {2018}{\natexlab{b}})\BibitemShut {NoStop}%
\bibitem [{\citenamefont {Okada}\ \emph {et~al.}(1991)\citenamefont {Okada}, \citenamefont {Katayama},\ and\ \citenamefont {Tominaga}}]{Okada1991}%
  \BibitemOpen
  \bibfield  {author} {\bibinfo {author} {\bibfnamefont {M.}~\bibnamefont {Okada}}, \bibinfo {author} {\bibfnamefont {S.}~\bibnamefont {Katayama}},\ and\ \bibinfo {author} {\bibfnamefont {K.}~\bibnamefont {Tominaga}},\ }\href {https://doi.org/10.1063/1.348498} {\bibfield  {journal} {\bibinfo  {journal} {Journal of Applied Physics}\ }\textbf {\bibinfo {volume} {69}},\ \bibinfo {pages} {3566} (\bibinfo {year} {1991})}\BibitemShut {NoStop}%
\bibitem [{\citenamefont {Sun}\ \emph {et~al.}(2013)\citenamefont {Sun}, \citenamefont {Chang}, \citenamefont {Kabatek}, \citenamefont {Song}, \citenamefont {Wang}, \citenamefont {Jantz}, \citenamefont {Schneider}, \citenamefont {Wu}, \citenamefont {Montoya}, \citenamefont {Kardasz}, \citenamefont {Heinrich}, \citenamefont {te~Velthuis}, \citenamefont {Schultheiss},\ and\ \citenamefont {Hoffmann}}]{Sun2013}%
  \BibitemOpen
  \bibfield  {author} {\bibinfo {author} {\bibfnamefont {Y.}~\bibnamefont {Sun}}, \bibinfo {author} {\bibfnamefont {H.}~\bibnamefont {Chang}}, \bibinfo {author} {\bibfnamefont {M.}~\bibnamefont {Kabatek}}, \bibinfo {author} {\bibfnamefont {Y.-Y.}\ \bibnamefont {Song}}, \bibinfo {author} {\bibfnamefont {Z.}~\bibnamefont {Wang}}, \bibinfo {author} {\bibfnamefont {M.}~\bibnamefont {Jantz}}, \bibinfo {author} {\bibfnamefont {W.}~\bibnamefont {Schneider}}, \bibinfo {author} {\bibfnamefont {M.}~\bibnamefont {Wu}}, \bibinfo {author} {\bibfnamefont {E.}~\bibnamefont {Montoya}}, \bibinfo {author} {\bibfnamefont {B.}~\bibnamefont {Kardasz}}, \bibinfo {author} {\bibfnamefont {B.}~\bibnamefont {Heinrich}}, \bibinfo {author} {\bibfnamefont {S.~G.~E.}\ \bibnamefont {te~Velthuis}}, \bibinfo {author} {\bibfnamefont {H.}~\bibnamefont {Schultheiss}},\ and\ \bibinfo {author} {\bibfnamefont {A.}~\bibnamefont {Hoffmann}},\ }\href {https://doi.org/10.1103/PhysRevLett.111.106601} {\bibfield  {journal} {\bibinfo  {journal} {Physical
  Review Letters}\ }\textbf {\bibinfo {volume} {111}},\ \bibinfo {pages} {106601} (\bibinfo {year} {2013})}\BibitemShut {NoStop}%
\bibitem [{\citenamefont {Bhoi}\ \emph {et~al.}(2018)\citenamefont {Bhoi}, \citenamefont {Kim}, \citenamefont {Kim}, \citenamefont {Kim}, \citenamefont {Lee},\ and\ \citenamefont {Kim}}]{Bhoi2018}%
  \BibitemOpen
  \bibfield  {author} {\bibinfo {author} {\bibfnamefont {B.}~\bibnamefont {Bhoi}}, \bibinfo {author} {\bibfnamefont {B.}~\bibnamefont {Kim}}, \bibinfo {author} {\bibfnamefont {Y.}~\bibnamefont {Kim}}, \bibinfo {author} {\bibfnamefont {M.-K.}\ \bibnamefont {Kim}}, \bibinfo {author} {\bibfnamefont {J.-H.}\ \bibnamefont {Lee}},\ and\ \bibinfo {author} {\bibfnamefont {S.-K.}\ \bibnamefont {Kim}},\ }\bibfield  {journal} {\bibinfo  {journal} {Journal of Applied Physics}\ }\textbf {\bibinfo {volume} {123}},\ \href {https://doi.org/10.1063/1.5031198} {10.1063/1.5031198} (\bibinfo {year} {2018})\BibitemShut {NoStop}%
\bibitem [{\citenamefont {Zhang}\ \emph {et~al.}(2017)\citenamefont {Zhang}, \citenamefont {Song},\ and\ \citenamefont {Chai}}]{Zhang2017}%
  \BibitemOpen
  \bibfield  {author} {\bibinfo {author} {\bibfnamefont {D.}~\bibnamefont {Zhang}}, \bibinfo {author} {\bibfnamefont {W.}~\bibnamefont {Song}},\ and\ \bibinfo {author} {\bibfnamefont {G.}~\bibnamefont {Chai}},\ }\href {https://doi.org/10.1088/1361-6463/aa68cf} {\bibfield  {journal} {\bibinfo  {journal} {Journal of Physics D: Applied Physics}\ }\textbf {\bibinfo {volume} {50}},\ \bibinfo {pages} {205003} (\bibinfo {year} {2017})}\BibitemShut {NoStop}%
\bibitem [{\citenamefont {Bhoi}\ \emph {et~al.}(2013)\citenamefont {Bhoi}, \citenamefont {Venkataramani}, \citenamefont {Aiyar},\ and\ \citenamefont {Prasad}}]{Bhoi2013}%
  \BibitemOpen
  \bibfield  {author} {\bibinfo {author} {\bibfnamefont {B.}~\bibnamefont {Bhoi}}, \bibinfo {author} {\bibfnamefont {N.}~\bibnamefont {Venkataramani}}, \bibinfo {author} {\bibfnamefont {R.~P. R.~C.}\ \bibnamefont {Aiyar}},\ and\ \bibinfo {author} {\bibfnamefont {S.}~\bibnamefont {Prasad}},\ }\href {https://doi.org/10.1109/TMAG.2012.2228172} {\bibfield  {journal} {\bibinfo  {journal} {IEEE Transactions on Magnetics}\ }\textbf {\bibinfo {volume} {49}},\ \bibinfo {pages} {990} (\bibinfo {year} {2013})}\BibitemShut {NoStop}%
\bibitem [{\citenamefont {Jermain}\ \emph {et~al.}(2017)\citenamefont {Jermain}, \citenamefont {Aradhya}, \citenamefont {Reynolds}, \citenamefont {Buhrman}, \citenamefont {Brangham}, \citenamefont {Page}, \citenamefont {Hammel}, \citenamefont {Yang},\ and\ \citenamefont {Ralph}}]{Jermain2017}%
  \BibitemOpen
  \bibfield  {author} {\bibinfo {author} {\bibfnamefont {C.~L.}\ \bibnamefont {Jermain}}, \bibinfo {author} {\bibfnamefont {S.~V.}\ \bibnamefont {Aradhya}}, \bibinfo {author} {\bibfnamefont {N.~D.}\ \bibnamefont {Reynolds}}, \bibinfo {author} {\bibfnamefont {R.~A.}\ \bibnamefont {Buhrman}}, \bibinfo {author} {\bibfnamefont {J.~T.}\ \bibnamefont {Brangham}}, \bibinfo {author} {\bibfnamefont {M.~R.}\ \bibnamefont {Page}}, \bibinfo {author} {\bibfnamefont {P.~C.}\ \bibnamefont {Hammel}}, \bibinfo {author} {\bibfnamefont {F.~Y.}\ \bibnamefont {Yang}},\ and\ \bibinfo {author} {\bibfnamefont {D.~C.}\ \bibnamefont {Ralph}},\ }\href {https://doi.org/10.1103/PhysRevB.95.174411} {\bibfield  {journal} {\bibinfo  {journal} {Physical Review B}\ }\textbf {\bibinfo {volume} {95}},\ \bibinfo {pages} {174411} (\bibinfo {year} {2017})}\BibitemShut {NoStop}%
\bibitem [{\citenamefont {Hyde}\ \emph {et~al.}(2016)\citenamefont {Hyde}, \citenamefont {Bai}, \citenamefont {Harder}, \citenamefont {Match},\ and\ \citenamefont {Hu}}]{Hyde2016}%
  \BibitemOpen
  \bibfield  {author} {\bibinfo {author} {\bibfnamefont {P.}~\bibnamefont {Hyde}}, \bibinfo {author} {\bibfnamefont {L.}~\bibnamefont {Bai}}, \bibinfo {author} {\bibfnamefont {M.}~\bibnamefont {Harder}}, \bibinfo {author} {\bibfnamefont {C.}~\bibnamefont {Match}},\ and\ \bibinfo {author} {\bibfnamefont {C.-M.}\ \bibnamefont {Hu}},\ }\href {https://doi.org/10.1063/1.4964602} {\bibfield  {journal} {\bibinfo  {journal} {Applied Physics Letters}\ }\textbf {\bibinfo {volume} {109}},\ \bibinfo {pages} {152405} (\bibinfo {year} {2016})}\BibitemShut {NoStop}%
\bibitem [{\citenamefont {Kosen}\ \emph {et~al.}(2019)\citenamefont {Kosen}, \citenamefont {van Loo}, \citenamefont {Bozhko}, \citenamefont {Mihalceanu},\ and\ \citenamefont {Karenowska}}]{Kosen2019}%
  \BibitemOpen
  \bibfield  {author} {\bibinfo {author} {\bibfnamefont {S.}~\bibnamefont {Kosen}}, \bibinfo {author} {\bibfnamefont {A.~F.}\ \bibnamefont {van Loo}}, \bibinfo {author} {\bibfnamefont {D.~A.}\ \bibnamefont {Bozhko}}, \bibinfo {author} {\bibfnamefont {L.}~\bibnamefont {Mihalceanu}},\ and\ \bibinfo {author} {\bibfnamefont {A.~D.}\ \bibnamefont {Karenowska}},\ }\href {https://doi.org/10.1063/1.5115266} {\bibfield  {journal} {\bibinfo  {journal} {APL Materials}\ }\textbf {\bibinfo {volume} {7}},\ \bibinfo {pages} {101120} (\bibinfo {year} {2019})}\BibitemShut {NoStop}%
\bibitem [{\citenamefont {Haidar}\ \emph {et~al.}(2016)\citenamefont {Haidar}, \citenamefont {Dürrenfeld}, \citenamefont {Ranjbar}, \citenamefont {Balinsky}, \citenamefont {Fazlali}, \citenamefont {Dvornik}, \citenamefont {Dumas}, \citenamefont {Khartsev},\ and\ \citenamefont {Åkerman}}]{Haidar2016}%
  \BibitemOpen
  \bibfield  {author} {\bibinfo {author} {\bibfnamefont {M.}~\bibnamefont {Haidar}}, \bibinfo {author} {\bibfnamefont {P.}~\bibnamefont {Dürrenfeld}}, \bibinfo {author} {\bibfnamefont {M.}~\bibnamefont {Ranjbar}}, \bibinfo {author} {\bibfnamefont {M.}~\bibnamefont {Balinsky}}, \bibinfo {author} {\bibfnamefont {M.}~\bibnamefont {Fazlali}}, \bibinfo {author} {\bibfnamefont {M.}~\bibnamefont {Dvornik}}, \bibinfo {author} {\bibfnamefont {R.~K.}\ \bibnamefont {Dumas}}, \bibinfo {author} {\bibfnamefont {S.}~\bibnamefont {Khartsev}},\ and\ \bibinfo {author} {\bibfnamefont {J.}~\bibnamefont {Åkerman}},\ }\href {https://doi.org/10.1103/PhysRevB.94.180409} {\bibfield  {journal} {\bibinfo  {journal} {Physical Review B}\ }\textbf {\bibinfo {volume} {94}},\ \bibinfo {pages} {180409} (\bibinfo {year} {2016})}\BibitemShut {NoStop}%
\end{thebibliography}%

\end{document}